\documentclass[aps,prb,amsmath,amssymb,reprint,
                letterpaper,footinbib,raggedbottom,%
                superscriptaddress,citeautoscript,balancelastpage,%
                floatfix]{revtex4-1}

\usepackage[usenames,dvipsnames]{xcolor}
\usepackage[bookmarks=false,colorlinks]{hyperref}
\hypersetup{
    linkcolor=magenta,        
    citecolor=PineGreen,     
    filecolor=Plum,         
    urlcolor=blue,           
}
\usepackage{graphicx}
\usepackage{mdwlist}
\usepackage[protrusion=true,expansion=true,final]{microtype}



\begin{document}

\title{Crystal Structure and Electronic Properties of Bulk and Thin Film Brownmillerite Oxides}
\author{Joshua Young}
	\email{jy346@drexel.edu}
	\affiliation{Department of Materials Science and Engineering, Drexel University, Philadelphia, PA 19104, USA}
  \affiliation{Department of  Materials  Science  and  Engineering,  Northwestern  University,  Evanston,  IL  60208,  USA}
\author{James M.\ Rondinelli}
	\email{jrondinelli@northwestern.edu}
  \affiliation{Department of  Materials  Science  and  Engineering,  Northwestern  University,  Evanston,  IL  60208,  USA}
	\affiliation{Materials Science Division, Argonne National Laboratory, Argonne, Illinois 60439, USA}
\date{\today}

\begin{abstract}
The equilibrium structure and functional properties exhibited by brownmillerite oxides, a family of perovskite-derived structures with alternating layers of $B$O$_6$ octahedra and $B$O$_4$ tetrahedra, \emph{viz}., ordered arrangements of oxygen vacancies, is dependent on a variety of competing crystal-chemistry factors. We use electronic structure calculations to disentangle the complex interactions in two ferrates, Sr$_2$Fe$_2$O$_5$ and Ca$_2$Fe$_2$O$_5$, relating the stability of the equilibrium (strain-free) and thin film structures to both previously identified and newly herein proposed descriptors. We show that cation size and intralayer separation of the tetrahedral chains provide key contributions to the preferred ground state. We show the bulk ground state structure is retained in the ferrates over a range of strain values; however, a change in the orientation of the tetrahedral chains, \emph{i.e.}, a perpendicular orientation of the vacancies relative to the substrate, is stabilized in the compressive region. The structure stability under strain is largely governed by maximizing the intraplane separation of the `dipoles' generated from rotations of the FeO$_4$ tetrahedra. Lastly, we find that the electronic band gap is strongly influenced by strain, manifesting as an unanticipated asymmetric-vacancy alignment dependent response. This atomistic understanding establishes a practical route for the design of novel functional electronic materials in thin film geometries.
\end{abstract}

\maketitle

\section{Introduction}

\begin{figure}
\centering
\includegraphics[width=0.99\columnwidth,clip]{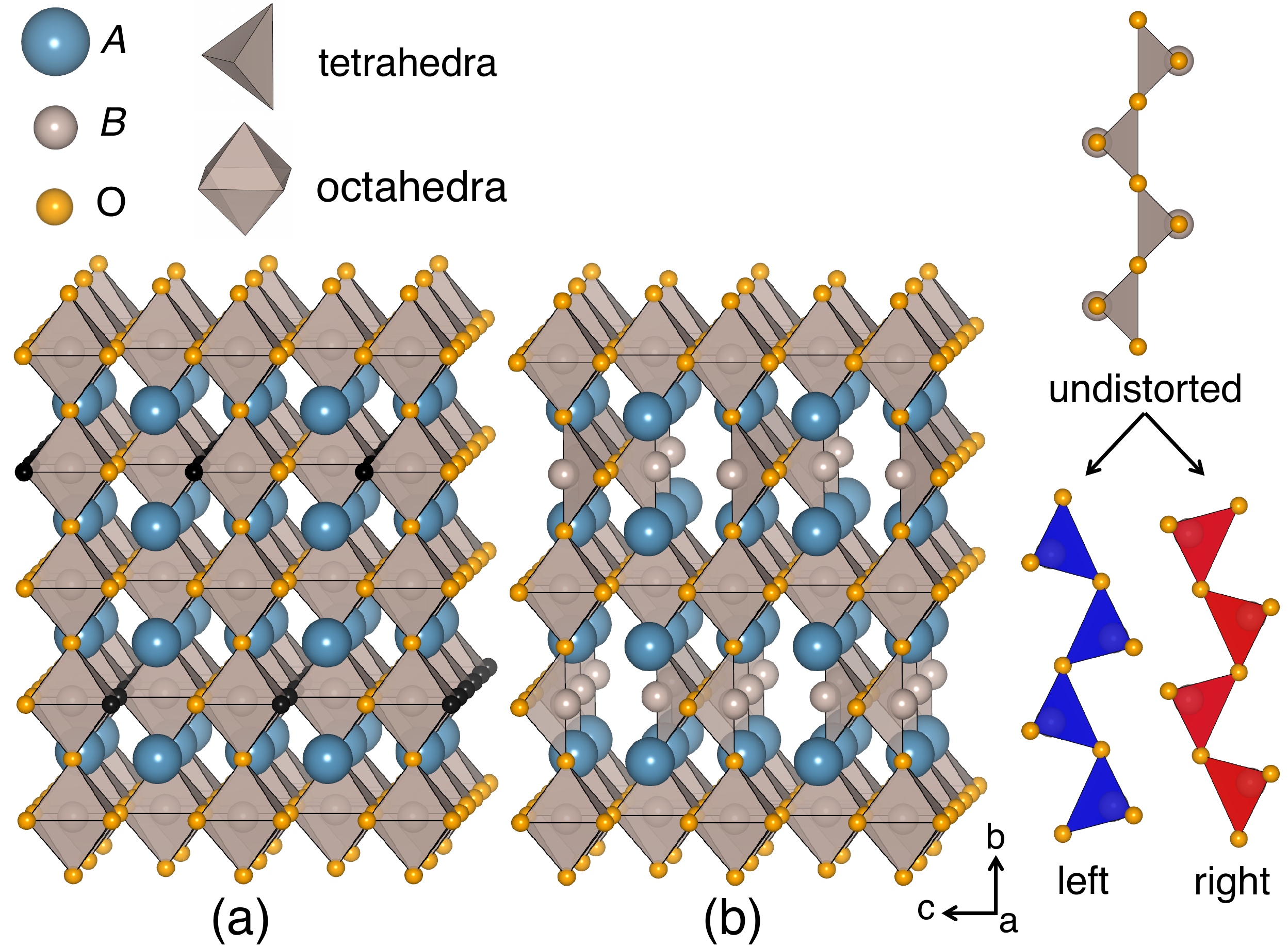}
\caption{The removal of chains of oxygen atoms (colored in black) from the $AB$O$_3$ perovskite structure (a) results in the vacancy ordered $AB$O$_{2.5}$ brownmillerite structure (b), which consists of alternating layers of $B$O$_4$ tetrahedra and $B$O$_6$ octahedral along the $b$ direction. The undistorted rows of $B$O$_4$ tetrahedra can then `twist' to create left- or right-handed chains (colored blue and red, respectively).
}
\label{fig:bulk}
\end{figure}

The family of brownmillerite oxides (general formula $AB$O$_{2.5}$ or $A_2$$B_2$O$_5$) are highly studied for use in ionic conducting and anion insertion applications.\cite{Kendall_etal:1995,Boivin/Mairesse:1998,Rolle_etal:2005,Shaula_etal:2006,Orera/Slater:2010,Antipov_etal:2004,Sullivan/Greaves:2012,Tarasova_etal:2013} 
Their structure type can be thought of as an $AB$O$_3$ perovskite with one-sixth of the oxygen atoms removed, creating parallel rows of ordered anion vacancies along the [110] crystallographic direction; this results in alternating layers of corner-connected $B$O$_4$ tetrahedra and $B$O$_6$ octahedra (Figure \ref{fig:bulk}a).
While cooperative distortions and octahedral rotations are well understood in perovskites, the presence of tetrahedral layers in the brownmillerites adds additional structural complexity and possible degrees of freedom for both structure and electronic function design.
In addition to rotations of the octahedra, each tetrahedral chain can `twist' in a ``left-handed" or ``right-handed" sense, resulting in two different types of chains related by mirror symmetry (Figure \ref{fig:bulk}b).
Furthermore, these different chains can be ordered relative to each other within the brownmillerite unit cell, 
resulting in a variety of structures displaying different space group symmetries.

\begin{figure}
\centering
\includegraphics[width=0.99\columnwidth,clip]{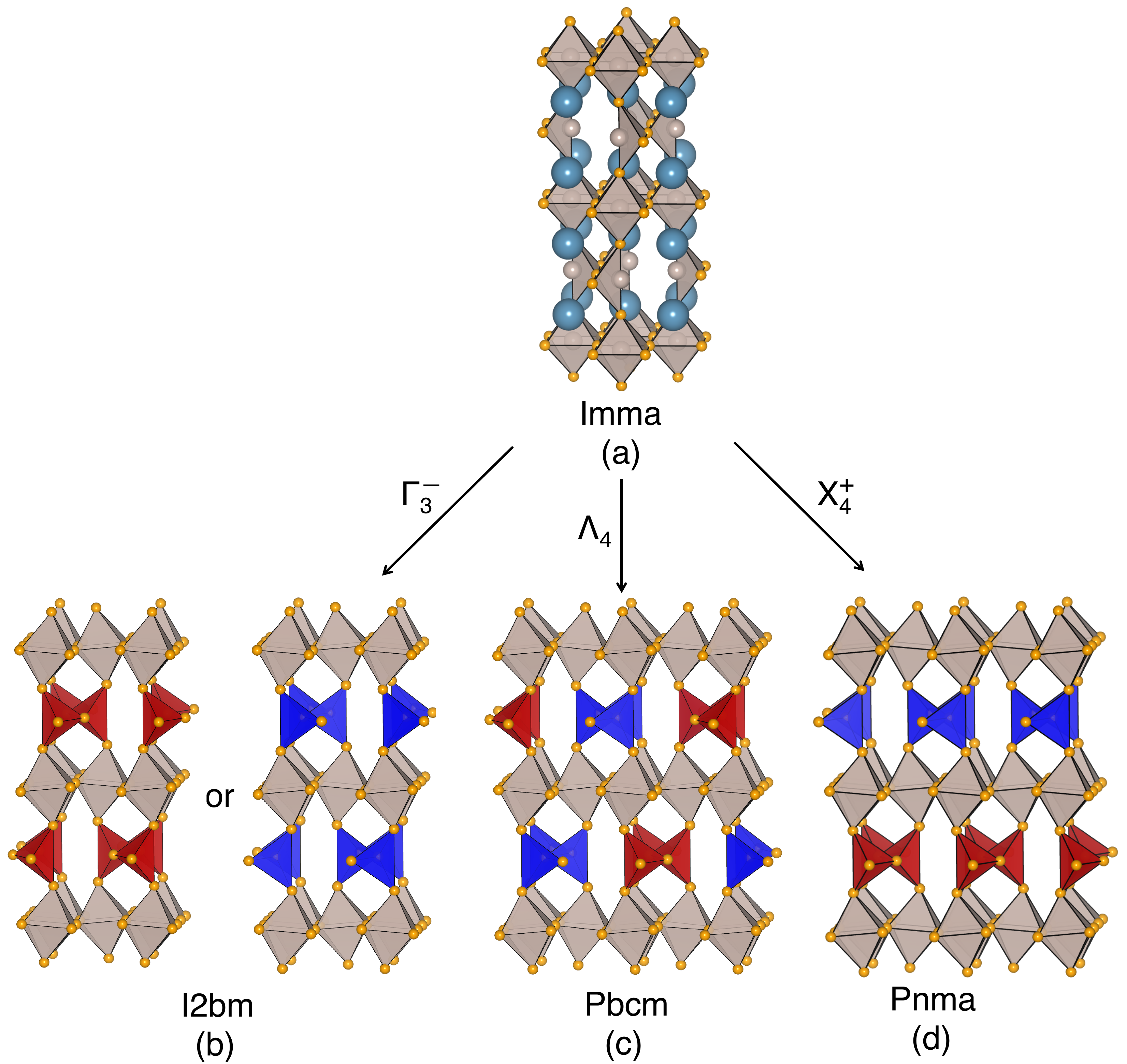}
\caption{The hypothetical high symmetry brownmillerite structure (a) is defined as having no octahedral or tetrahedral rotations, and has the $Imma$ space group. Relative ordering of tetrahedral chains results in three different low-symmetry structures. If all chains are of the same handedness, the structure is polar $I2bm$ (b); alternation of left- and right-handed chains within each layer results in centric $Pbcm$ (c), while alternation between each layer gives centric $Pnma$ (d). The $A$-site cations are omitted from the low-symmetry structures for clarity.
}
\label{fig:tet_structures}
\end{figure}

When the tetrahedra and octahedra are not rotated, the structure displays the $Imma$ space group (Figure \ref{fig:tet_structures}a); this aristotype may be used as a 
high symmetry reference phase for which subsequent structural analyses are made.
Alternatively, the $Imma$ structure can be imagined as having disordered left- and right-handed chains (\textit{i.e.} the chains display either incommensurate ordering or no long range order); some structures, such as Sr$_2$CoFeO$_5$ and Sr$_2$MnGeO$_5$,\cite{Battle_etal:1988,Abakumov_etal:2003} display this phase at ambient conditions, while others become disordered at high temperature (including Ca$_2$Al$_2$O$_5$ and Ca$_2$Fe$_2$O$_5$).\cite{Lazic_etal:2008,Kruger/Kahlenberg:2005}
Each of the three low-symmetry bulk hettotypes display two-dimensional sheets of corner-connected octahedra, which rotate out-of-phase along the $ac$ direction. This rotation transforms like the irreducible representation (irrep) $\Gamma_1^+$ of the $Imma$ phase.

Along the $b$ axis, however, the allowed displacements of the oxygen atoms are constrained by the `handedness' and ordering of the tetrahedral chains, 
which ultimately control the final symmetry of the brownmillerite structure (see the discussion in Section \ref{sec:discussion}).
The tetrahedral chains can cooperatively rotate in a variety of ways, each described by a different irrep of $Imma$.
When the tetrahedra rotate into either all left- or all right-handed chains the structure displays the polar space group $I2bm$ owing to the 
$\Gamma_3^-$ irrep (Figure \ref{fig:tet_structures}b).
If there is a racemic mixture of both types of chains, the structure becomes centrosymmetric, with different relative orderings generating different symmetries.
Alternating chains of different handedness within each tetrahedral layer is described by the $\Lambda_4$ irrep and yields the 
centrosymmetric $Pbcm$ structure (Figure \ref{fig:tet_structures}c), while alternation between layers (given by $X_4^+$) gives the 
centric $Pnma$ structure (Figure \ref{fig:tet_structures}d).
Because the left- and right-handed chains are related by symmetry and differ only by small atomic displacements, the formation energies for the different 
polymorphs are nearly degenerate and should form with equal probability.
However, each of the aforementioned ordering types is seen experimentally in various members of the brownmillerite family, and the driving force behind the preferred type in different chemistries is not completely understood.

\begin{figure}
\centering
\includegraphics[width=0.99\columnwidth,clip]{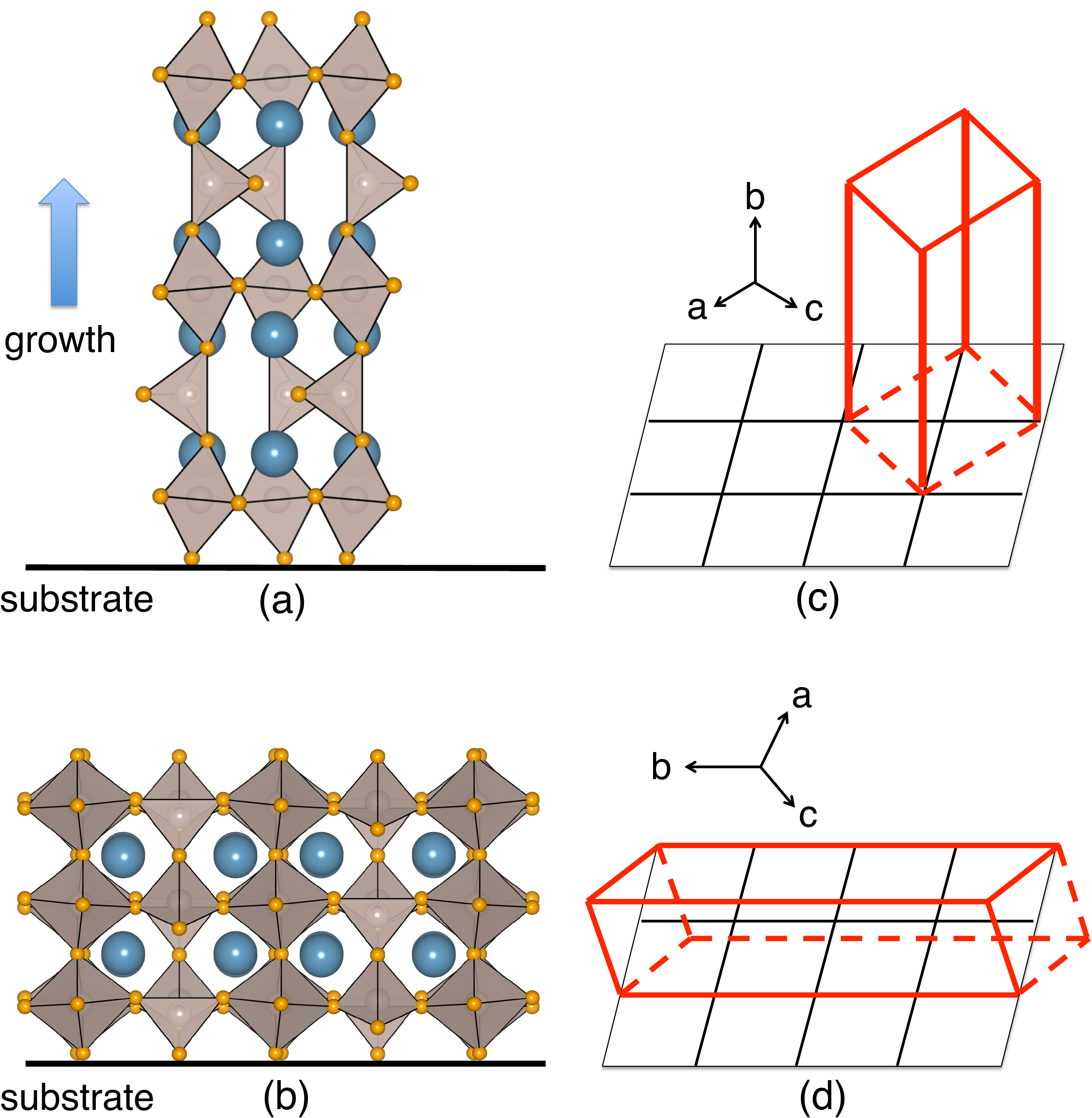}
\caption{When brownmillerite structures are placed under epitaxial strain, the oxygen deficient layers can order (a) parallel or (b) perpendicular to the substrate, with the pseudocubic orientations shown in (c) and (d), respectively.}
\label{fig:substrate}
\end{figure}

Another degree of freedom appears for the brownmillerite structures when 
when they are grown as a thin film, \emph{e.g.}, via molecular beam epitaxy or pulsed laser deposition, 
owing to the constraints imposed by epitaxial strain.
In this thin film case, the oxygen-deficient layers can order parallel or perpendicular to the substrate (this is shown in Figure \ref{fig:substrate}a and \ref{fig:substrate}b, respectively, with the pseudocubic orientation shown in Figure \ref{fig:substrate}c and \ref{fig:substrate}d).
Although different strain states will stabilize one orientation over the other, it is not always clear which will be preferred and why.
In (La,Sr)Co$_2$O$_5$, for example, tensile strain stabilizes perpendicular ordering while compressive strain stabilizes parallel ordering;\cite{Klenov_etal:2003,Gazquez_etal:2013} however, the opposite effect is observed in strained Ca$_2$Fe$_2$O$_5$.\cite{Inoue_etal:2010}

In this work, we seek to disentangle the structural and energetic effects operative in brownmillerite oxides 
through an investigation of bulk and epitaxially strained Sr$_2$Fe$_2$O$_5$ and Ca$_2$Fe$_2$O$_5$ using first-principles density functional theory calculations.
The Sr$_2$Fe$_2$O$_5$ and Ca$_2$Fe$_2$O$_5$ members are experimentally known to be stable in the ordered brownmillerite phase up to high temperatures (approximately 1200 K for Sr$_2$Fe$_2$O$_5$ and 920 K for Ca$_2$Fe$_2$O$_5$);\cite{Schmidt/Campbell:2001,Kruger_2009} at ambient conditions, Sr$_2$Fe$_2$O$_5$ displays the \textit{Pbcm} ordering,\cite{Auckett_etal:2012,Auckett_etal:2013} while Ca$_2$Fe$_2$O$_5$ exhibits the \textit{Pnma} structure.\cite{Kruger_2009}
Additionally, they both contain only Fe$^{3+}$ cations and display G-type antiferromagnetic ordering with high 
N\'{e}el temperatures (700 and 720 K, respectively).\cite{Takeda_etal_2:1968,Takeda_etal_1:1969,Schmidt/Campbell:2001}
The indirect band gap of both compounds is approximately 2.0 eV, owing to a $\Gamma$- to X-point transition.\cite{Galakhov_etal:2010}

We find that, in agreement with experimental results of the bulk phases, the \textit{Pbcm} and \textit{Pnma} phase are the lowest energy (and thus equilibrium) structures for Sr$_2$Fe$_2$O$_5$ and Ca$_2$Fe$_2$O$_5$, respectively.
Furthermore, we find that the parallel (or perpendicular) ordering is preferred under tensile (or compressive) strain; additionally, this switch occurs due to which ordering maximizes the intralayer tetrahedral chain distance under a specific strain state, analogously to the bulk phases.
Finally, we report that the band gap of these materials is heavily influenced by epitaxial strain owing to 
strain-induced changes to the intralayer bond angles.
Although strain is well known to influence the band gap via control 
of the octahedral rotations in $AB$O$_3$ perovskites, the alternating 
octahedra-tetrahedra layers in the brownmillerites allow for a more complex coupling between the two.
We show that, similar to the $AB$O$_3$ perovskites, an alteration of the $B$--O--$B$ bond angles causes this response; unlike perovskites, however, it is the angle between the tetrahedrally and octahedrally coordinated irons which controls the band gap.
The band gap of the compounds in the parallel orientation increases under increasing tensile strain, but decreases in the perpendicular orientation, a discrepancy that arises from how the response of the out-of-plane lattice parameter to strain affects this bond angle differently in the two orientations. 

\begin{figure}
\centering
\includegraphics[width=0.99\columnwidth,clip]{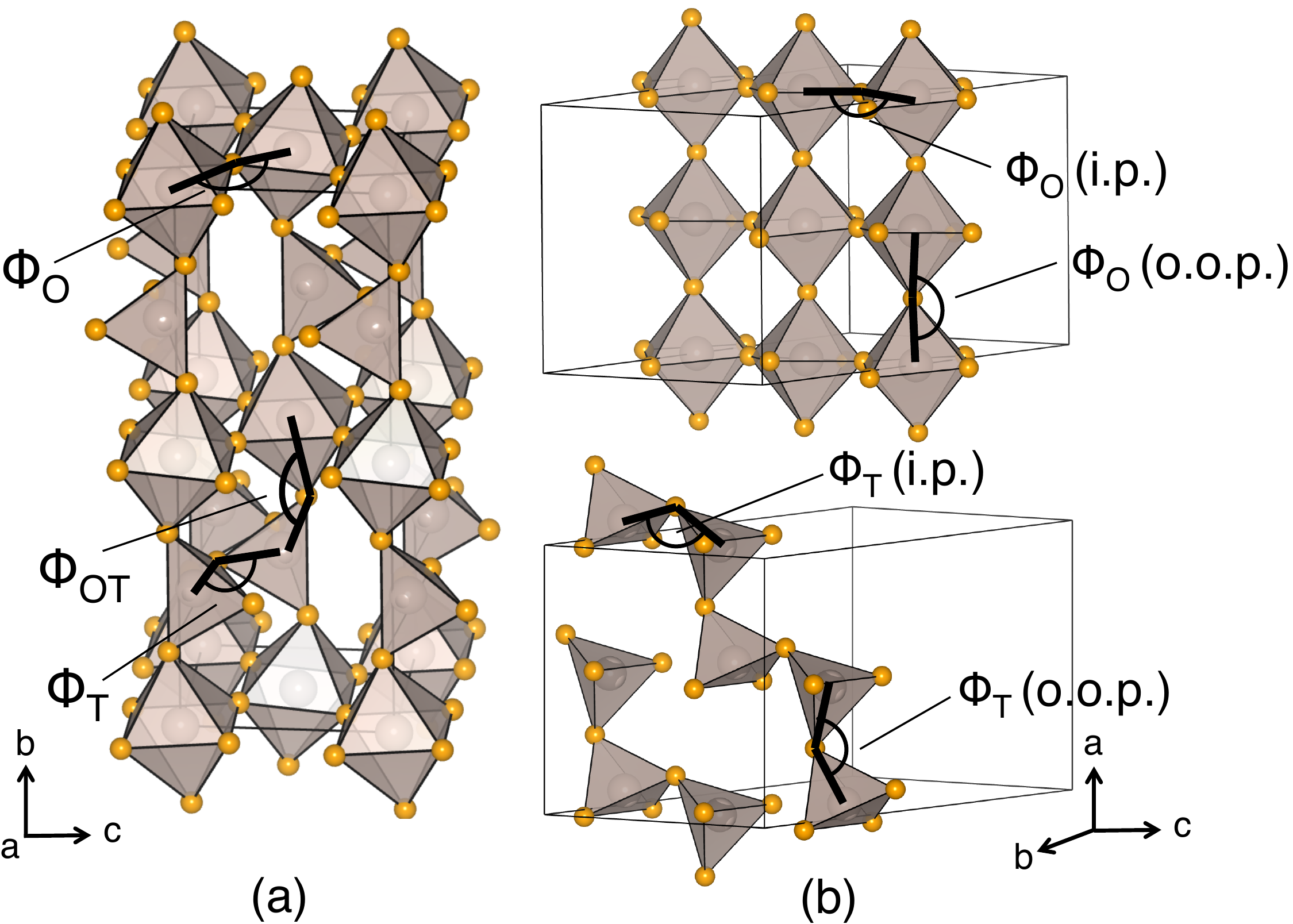}
\caption{The three characteristic angles of the $A_2B_2$O$_5$ brownmillerite structure (a). Rotations of the tetrahedral chains, of the octahedral network, and between the tetrahedral and octahedral layers are given by $\Phi_T$,  $\Phi_O$, and  $\Phi_{OT}$, respectively. These three angles are also sufficient to describe the case when the tetrahedral chains order parallel to the substrate under biaxial strain. In the perpendicular orientation (b), however, the changes in the out-of-plane lattice parameter owing to the strain state splits the octahedral and tetrahedral rotation angles.
}
\label{fig:angles_bulk}
\end{figure}

\section{Computational Methods}

All investigations were performed using density functional theory as implemented in the Vienna \textit{ab-initio} Simulation Package (\texttt{VASP}).\cite{Hohenberg/Kohn:1964,Kresse/Hafner:1993,Kresse/Hafner:1996}
We used projector augmented-wave (PAW) potentials with the PBEsol functional, with valence electron configurations of $3s^23p^64s^2$ for Ca, $4s^24p^65s^2$ for Sr, $3d^74s^1$ for Fe, and $2s^22p^4$ for O.\cite{Blochl:1994,PBEsol:2008}
A plane-wave cutoff of 500 eV and a 7$\times$5$\times$7 Monkhorst-Pack mesh was used during the structural relaxations.\cite{Monkhorst/Pack:1976}
We applied a Hubbard $U$ correction of 5 eV using the Dudarev formalism to treat the correlated Fe $3d$ states; we also enforced a G-type anti-ferromagnetic collinear spin ordering on the Fe atoms.\cite{Dudarev:1998}
Symmetry-adapted mode decompositions were performed using the ISODISTORT tool, part of the ISOTROPY software suite.\cite{ISODISTORT}
Atomic structures were visualized using VESTA.\cite{VESTA}

To simulate the application of epitaxial strain by growth on a cubic [001] terminated perovskite substrate with a square surface net, we fix the $a$ and $c$ lattice parameters to be equal and allow the out-of-plane $b$ axis and ions to fully relax (adopted from the approach of Ref. \citenum{Zayak_etal:2006}).
The $I2bm$, $Pbcm$, and $Pnma$ phases of Ca$_2$Fe$_2$O$_5$ and the $I2bm$ and $Pbcm$ phases of Sr$_2$Fe$_2$O$_5$, in both the parallel and perpendicular orientation were strained from -3\% to 3\% in increments of 1\%. 
Because of the difference in size and equilibrium volume of the Ca$_2$Fe$_2$O$_5$ and Sr$_2$Fe$_2$O$_5$ unit cells, application of the same percent strain results in different pseudocubic lattice parameters (a$_{pc}$) for the two structures; for this reason, we report the strain in terms of $a_{pc}$ rather than in terms of percentage.

\section{\label{sec:discussion}Results and Discussion}

\subsection{Polyhedral Rotation Definitions}
To characterize the brownmillerite structures, we consider three types of rotation angles: the angle between (i)  the FeO$_4$ tetrahedra ($\Theta_{T}$), (ii) the FeO$_6$ octahedra ($\Theta_{O}$), and (iii) the tetrahedra and octahedra ($\Theta_{OT}$).
We report these angles as $\Theta_{X} = (180 - \Phi_{X})/2$, where the definition of $\Phi_X$ ($X$ = $T$, $O$, or $OT$) is shown in Figure \ref{fig:angles_bulk}; in this case, a larger angle indicates a larger distortion of the relative bond away from 180$^{\circ}$.
In the bulk case, and the case of parallel orientation of vacancies under strain, these three angles are sufficient to describe the structure (Figure \ref{fig:angles_bulk}a).
In the perpendicular orientation, however, the applied strain and response of the out-of-plane lattice parameter now affect the tetrahedra and octahedra in different ways.
To capture this, we divide $\Phi_{O}$ and $\Phi_{T}$ into two unique angles defined by in-plane (i.p.) and out-of-plane (o.o.p.) components (Figure \ref{fig:angles_bulk}b).

\begingroup
\squeezetable
\begin{table*}[t]
\begin{ruledtabular}
\caption{\label{tab:bulk}The energetics and structure of bulk Sr$_2$Fe$_2$O$_5$ and Ca$_2$Fe$_2$O$_5$. The tolerance factor ($\tau$) is defined by Equation \ref{eq:tolfact} in the text. The energy difference between the different tetrahedral chain ordering structures is given as $\Delta E$; each energy is given as the difference in meV between that phase and the lowest energy phase for each compound normalized to the number of formula units. The $a$, $b$, and $c$ lattice parameters are given in Angstroms. The rotations of the tetrahedra, octahedra, and between the octahedra and tetrahedra are given in degrees as $\Theta_T$, $\Theta_O$, and $\Theta_{OT}$, respectively; each is defined in Figure \ref{fig:angles_bulk}. The average intralayer separation of tetrahedral chains is $R$ (defined in Figure \ref{fig:chaindistance}), and the deviation in the bond lengths of an octahedra is given by $\Delta$. The band gap ($E_g$) of each structure is given in eV.}
\begin{tabular}{lcccccccccc}
\\[-0.7em]
 & \multicolumn{10}{c}{Sr$_2$Fe$_2$O$_5$ ($P_T$ = 3.8 D, $\tau$=0.976)}  \\
 \hline
Symmetry & $\Delta E$ (meV/f.u.) & $a$ (\AA) & $b$ (\AA) & $c$ (\AA) & $\Theta_{T}$ ($^{\circ}$) & $\Theta_{O}$ ($^{\circ}$) & $\Theta_{OT}$ ($^{\circ}$) & $R$ (\AA) & $\Delta$ ($\times 10^{-4}$) & $E_g$ (eV)  \\
\hline
\textbf{\textit{I2bm}} & 12.6 & 5.501 & 15.402 & 5.659 & 24.48 & 3.70 & 15.41 & 5.0977 & 18.79 & 2.19  \\
$Pbcm$ & 0 & 5.503 & 15.407 & 11.311 & 24.54 & 4.43 & 15.39 & 5.0988 & 19.07 & 2.15 \\
$Pnma$ & 22.7 & 5.499 & 15.413 & 5.659 & 24.49 & 3.64 & 15.31 & 5.0971 & 19.03 & 2.09  \\
\hline\hline\\[-0.7em]
 & \multicolumn{10}{c}{Ca$_2$Fe$_2$O$_5$ ($P_T$ = 1.7 D, $\tau$=0.923)} \\\hline
Symmetry & $\Delta E$ (meV/f.u.) & $a$ (\AA) & $b$ (\AA) & $c$ (\AA) & $\Theta_{T}$ ($^{\circ}$) & $\Theta_{O}$ ($^{\circ}$) & $\Theta_{OT}$ ($^{\circ}$) & $R$ (\AA) & $\Delta$ ($\times 10^{-4}$) & $E_g$ (eV) \\
\hline
$I2bm$ & 16.7 & 5.381 & 14.617 & 5.579 & 27.92 & 6.538 & 20.34 & 5.0173 & 11.16 & 2.16  \\
$Pbcm$ & 23.3 & 5.386 & 14.627 & 11.155 & 28.16 & 4.909 & 19.90 & 5.0476 & 10.85 & 2.13  \\
\textbf{\textit{Pnma}} & 0 & 5.397 & 14.632 & 5.561 & 27.23 & 8.032 & 20.45 & 4.9941 & 10.73 & 2.07  \\
\end{tabular}
\end{ruledtabular}
\end{table*}
\endgroup

\subsection{Bulk Phases}

We first investigated the bulk phases of Sr$_2$Fe$_2$O$_5$ and Ca$_2$Fe$_2$O$_5$ with the three tetrahedral chain orderings shown in Figure \ref{fig:tet_structures}.
The main results are summarized in Table \ref{tab:bulk}.
In the case of Sr$_2$Fe$_2$O$_5$, we found that centrosymmetric $Pbcm$ is the lowest energy structure, followed by $I2bm$, with $Pnma$ being the highest.
In Ca$_2$Fe$_2$O$_5$, we found the $Pnma$ structure to be lowest in energy, followed by $I2bm$ and $Pbcm$.
Each polymorph is separated from the other two by a small amount of energy ($\Delta E$, Table \ref{tab:bulk}), indicating that the formation of a right- or left-handed tetrahedral chain may be equally probable.
Furthermore, the lattice parameters, rotation angles, and band gaps vary only slightly between orderings.
Between the two chemistries, Ca$_2$Fe$_2$O$_5$ has a smaller unit cell and larger rotation angles than Sr$_2$Fe$_2$O$_5$, which is due to the smaller size of the Ca$^{2+}$ cation.

Experimentally, the structure of Sr$_2$Fe$_2$O$_5$ has been highly contested.
Initial structure refinements on powder samples ambiguously supported assignment of both completely disordered tetrahedral chains (space group $Imma$) or pure left- or right-handed ordering (space group $I2bm$); this material was thus theorized to display $I2bm$ symmetry locally, but with random ordering taking place over longer length scales.
This was challenged, however, by transmission electron microscopy results showing clear intralayer alternation of tetrahedral chains.\cite{Dhondt_etal:2008}
Additionally, more recent neutron diffraction experiments on single crystals grown by the floating zone method have indicated that $Pbcm$ is indeed the preferred structure type for Sr$_2$Fe$_2$O$_5$. \cite{Auckett_etal:2012,Auckett_etal:2013}
Furthermore, all of these recent experiments are supported by the aforementioned results of our \textit{ab initio} calculations; the small energy difference between the $Pbcm$ and $I2bm$ structures, however, indicate that intergrowths of the phases is not out of the question.
In contrast to the ambiguity of Sr$_2$Fe$_2$O$_5$, the structure of Ca$_2$Fe$_2$O$_5$ is well known to form in the $Pnma$ phase,\cite{Kruger_2009} again in agreement with our theoretical results.

What factors lead to and stabilize the preferred ground state in different brownmillerite compounds?
In 2005, Abakumov \textit{et al.} put forth the idea that the twisting of tetrahedral chains away from the undistorted 180$^{\circ}$ orientation creates local dipole moments, with larger rotations producing larger dipoles.\cite{Abakumov_etal:2005}
Hadermann and Abakumov \textit{et al.} further suggested that the distance (\textit{i.e.}, the length of the $b$ axis) between the tetrahedral layers is also an important factor to consider.\cite{Hadermann/Abakumov_etal:2007}
Parsons \textit{et al.} then recognized that each tetrahedral ordering scheme distorts the octahedra in different ways; the fact that the octahedra are not connected out-of-plane (along $b$) causes the apical oxygen atoms to displace more than the equatorial ones, which creates a `shearing' effect from this non-rigid rotation.
Generally, the $I2bm$ phase causes the least octahedral (elastic) distortion, followed by $Pnma$, with $Pbcm$ causing the most.
They then rationalized these arguments into a ``structure map" relating the observed phase of different brownmillerites to these factors, followed by a classification of several known compounds into this scheme.\cite{Parsons_etal:2009}
Although these observations have been key in building an understanding of structural trends in brownmillerite oxides, and are corroborated by some recently synthesized brownmillerite phases (such as Ca$_2$Cr$_2$O$_5$),\cite{Lopez/Attfield_2015}discrepancies in this structure map show there are additional effects which should be considered.
Ca$_2$FeCoO$_5$ and Ca$_2$Co$_2$O$_5$, for example, both display the $Pbcm$ rather than the structure-map predicted $Pnma$ structure.\cite{Ramezanipour_etal:2010,Zhang/Mitchell_2014}

\begin{figure}
\centering
\includegraphics[width=0.75\columnwidth,clip]{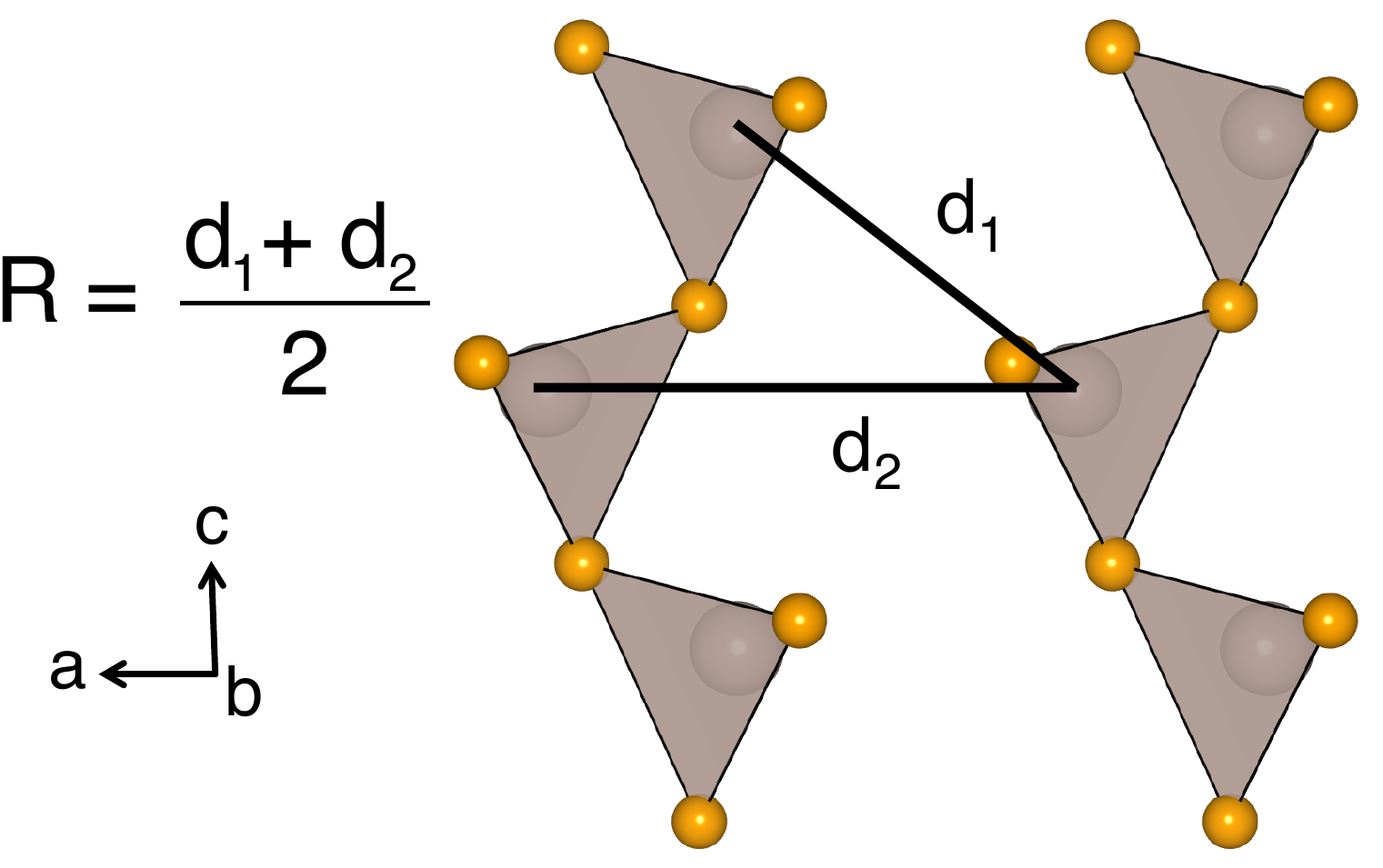}\vspace{-1\baselineskip}
\caption{The intralayer chain separation in the brownmillerite structure. The average separation distance is given by $R$, which is defined as the average of the two distances between Fe atoms of different chains ($d_1$ and $d_2$).
}
\label{fig:chaindistance}
\end{figure}

We now seek to explain the stability of these materials' preferred ground state phase in terms of local structure.
The tetrahedral chain ordering configuration preferred by the ground state can be summarized as a complex competition between two primary energetic factors: (\textit{i}) separation of the tetrahedral chains (\textit{i.e.}, minimization of electrostatic repulsion) and (\textit{ii}) distortion of the nominally regular $B$O$_6$ octahedra (minimization of elastic strain energy).
The three different ordering schemes shown in Figure \ref{fig:tet_structures} better maximize either factor (\textit{i}) or (\textit{ii}), at the expense of the other.
Following the approach of Zhang \textit{et al.},\cite{Zhang/Mitchell_2014} we can assign a magnitude to the local dipole generated by the tetrahedral rotations ($P_{T}$, Table \ref{tab:bulk}, given in units of Debye).
In compounds with large $P_T$, factor (\textit{i}) becomes the quantity of interest to maximize; rather than considering the separation along $b$, however, we quantify this by defining the average \textit{intra}layer separation of the tetrahedral chains as $R$ (shown in Figure \ref{fig:chaindistance}).
Competing with this is factor (\textit{ii}), \textit{i.e.}, the regularity in the octahedra.
We use the averaged sum-of-squares difference between the measured bond lengths ($d_n$) and the average bond length ($d_{avg}$) in the octahedra to quantify this:
\begin{equation}
\label{eq:tolfact}
\Delta = \frac{1}{N}\sum\limits_{i=1}^n \left(\frac{d_n-d_{avg}}{d_{avg}}\right)^2.
\end{equation}
For small values of $P_T$, factor (\textit{ii}) becomes the critical quantity to optimize, and minimizing $\Delta$ becomes more important than maximizing $R$.

Interestingly, we found that although Sr$_2$Fe$_2$O$_5$ has smaller rotations, it has a larger chain dipole ($P_T$ = 3.8 D) than Ca$_2$Fe$_2$O$_5$ ($P_T$ = 1.7 D); generally, one would expect smaller rotations to generate a smaller dipole.
This is due to the fact that the electric polarization generated by the apical oxygen atoms tends to cancel that from the equatorial oxygen atoms in the tetrahedra; however, the smaller rotations of the \textit{octahedra} in Sr$_2$Fe$_2$O$_5$ means this cancellation happens to a lesser degree, increasing $P_T$.
Again, at large values of $P_T$, the need to separate the dipoles, factor (\textit{i}), overcomes the need to have regular octahedra, factor  (\textit{ii}).
Sr$_2$Fe$_2$O$_5$ therefore exhibits the $Pbcm$ structure, which allows for the best average in-plane separation ($R$, Table \ref{tab:bulk}) of the large dipoles.
In Ca$_2$Fe$_2$O$_5$, the larger octahedral rotations distort the octahedra more in the presence of tetrahedral chains; this, in combination with the small chain dipole, therefore gives more importance to the phase which keeps the octahedra most regular ($\Delta$, Table \ref{tab:bulk}).
In this case, that is the $Pnma$ structure.

Another approach to understanding these energetic competitions is to consider the the Goldschmidt tolerance factor.\cite{Goldschmidt:1926} 
This quantity is given by:
\begin{equation}
\label{eq:tolfact}
\tau = \frac{r_A + r_\mathrm{O}}{\sqrt{2}(r_B+r_\mathrm{O})},
\end{equation}
where $r_A$, $r_B$, and $r_\mathrm{O}$ is the radius of the $A$-site atom, $B$-site atom, and oxide anion, respectively.\footnote{Although this is generally intended for use in perovskites, it is also applicable to perovskite-derived structures such as these brownmillerites.}
An ideal undistorted (cubic or tetragonal) structure will have $\tau$=1; the further the tolerance factor deviates from 1, the more the structure is likely to distort, and the larger the octahedral rotations will be.
In Ca$_2$Fe$_2$O$_5$ ($\tau$=0.923), the $Pnma$ phase least distorts the octahedra (as reflected in $\Delta$ in Table \ref{tab:bulk}), and is therefore the most stable.
As $\tau$ increases, the need to maximize the separation between dipoles overcomes the steric packing driving force for octahedral rotations and hence they decrease; because $Pbcm$ best maximizes this distance ($R$ in Table \ref{tab:bulk}), it is the preferred phase for the larger Sr$_2$Fe$_2$O$_5$ ($\tau$=0.976).

Finally, despite the structural differences between the  bulk Sr$_2$Fe$_2$O$_5$ and Ca$_2$Fe$_2$O$_5$ phases, the DFT-PBEsol (indirect $\Gamma$-X) band gap remains nominally the same ($\sim$ 2.10\,eV) across changes in chemistry and tetrahedral ordering.
This is due to the fact that O $2p$ states make up the top of the valence band, while the bottom of the conduction band is made up of Fe $3d$ states, making the band gap largely independent of $A$-site chemistry.
A more detailed analysis of the changes in the electronic gap is given for the thin film cases below.

\begin{figure*}
\centering
\includegraphics[width=1.8\columnwidth,clip]{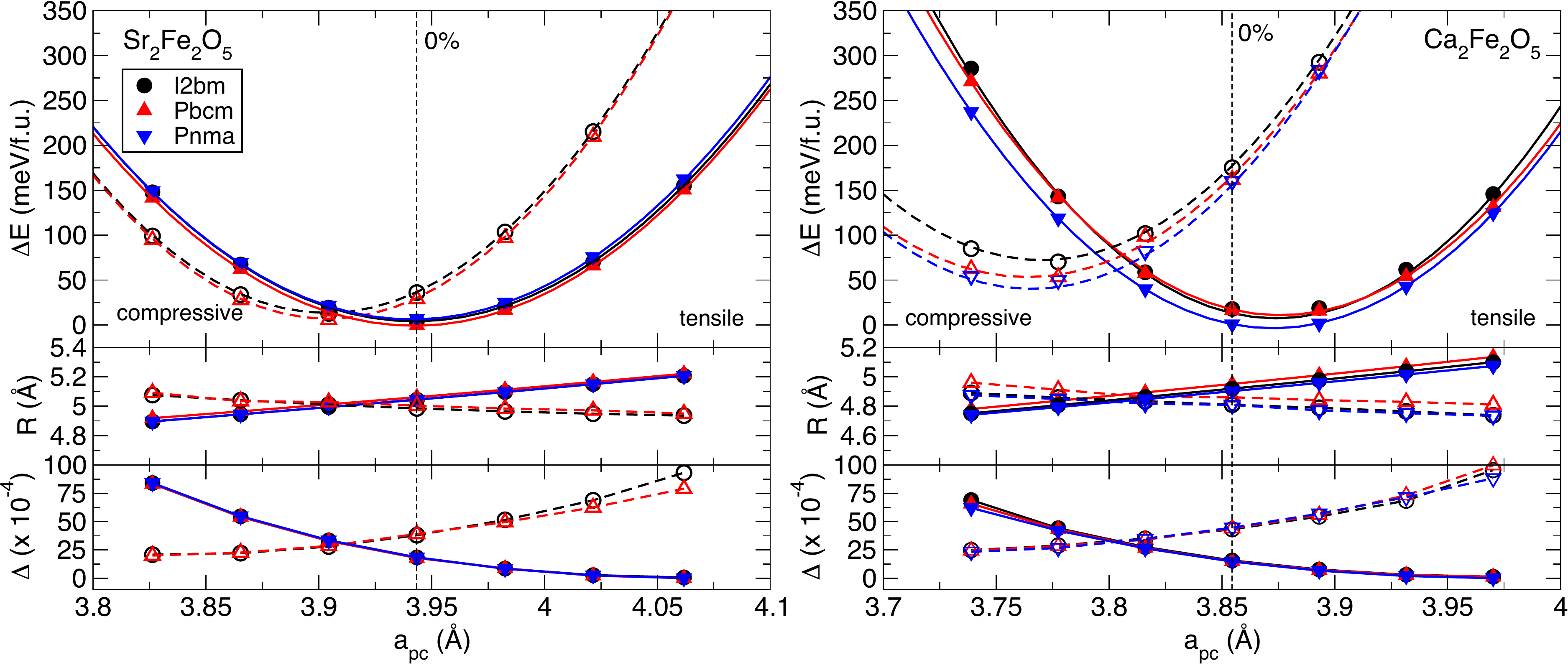}
\caption{Energy of the different tetrahedral chain ordered Sr$_2$Fe$_2$O$_5$ (left) and Ca$_2$Fe$_2$O$_5$ (right) structures as a function of epitaxial strain (top panels). In both cases, the parallel vacancy ordering (filled symbols) is stabilized under tensile strain, while perpendicular (empty symbols) is stabilized under compressive. This change in stabilization occurs at the point where either the parallel or perpendicular phase maximizes the average intralayer tetrahedral chain separation ($R$, middle panels) and minimizes the octahedral distortion effect ($\Delta$, bottom panels).
}
\label{fig:strainE}
\end{figure*}

\subsection{Strained Phases}

We next investigated the crystal and electronic structure of Sr$_2$Fe$_2$O$_5$ and Ca$_2$Fe$_2$O$_5$ under epitaxial strain.
As mentioned previously, we must now consider the orientation of the vacancies as another degree of freedom (\textit{i.e.}, whether the vacancy layers are parallel or perpendicular to the substrate, Figure \ref{fig:substrate}), in addition to tetrahedral chain rotation and ordering.
First, we find that when Sr$_2$Fe$_2$O$_5$ is constrained to a substrate, it retains the $Pbcm$ ground state in both parallel (filled symbols) and perpendicular  (empty symbols) substrate-vacancy orientations (Figure \ref{fig:strainE}a, top), a fact that can be explained using the same arguments presented for the bulk phase.
However, we also find that the perpendicular configuration of oxygen-deficient layers is stabilized over the parallel arrangement under small amounts of compressive strain.
This is due to the fact that increasing compressive strain \textit{decreases} the intralayer separation between tetrahedral chains if the vacancies are ordered parallel to the substrate (\textit{i.e.}, becomes more energetically unfavorable), but \textit{increases} $R$ if the vacancies are perpendicular.
The strain value at which the perpendicular ordering becomes more stable exactly coincides with the values at which the intralayer separation becomes greater (Figure \ref{fig:strainE}a, middle) and the octahedral distortions become smaller (Figure \ref{fig:strainE}a, bottom) than that of the parallel ordering.

One important detail to note is that Sr$_2$Fe$_2$O$_5$ has a relatively large pseudo-cubic lattice parameter ($a_{pc}$); only with some of the largest commercially available substrates can it be placed under tensile strain and have the parallel orientation be stabilized.
Many experimental observations agree with our prediction of the energetic stability of vacancy orientation.
%
%
Growth of Sr$_2$Fe$_2$O$_5$ near the transition point, such as on SrTiO$_3$ ($a_{pc}$ = 3.91 \AA) shows a competition between the two orientations,\cite{Shimakawa_etal:2010} whereas growth on larger substrates stabilizes the parallel orientation, such as on KTaO$_3$ ($a_{pc}$ = 3.99 \AA).\cite{Inoue_etal:2008}

When Ca$_2$Fe$_2$O$_5$ is placed under epitaxial strain, the $Pnma$ phase remains lowest in energy, while the $I2bm$ and $Pbcm$ structures become much closer in energy and strongly compete.
As with Sr$_2$Fe$_2$O$_5$, the tetrahedral layers in Ca$_2$Fe$_2$O$_5$ switch to a perpendicular orientation under compressive strain, and parallel under tensile (Figure \ref{fig:strainE}b, top).
Once again, this occurs due to the perpendicular orientation maximizing the distance $R$ between intralayer tetrahedral chains and minimizing octahedral distortions, $\Delta$, under compression (Figure \ref{fig:strainE}b, middle and bottom).
Although the $Pnma$ phase minimizes octahedral shearing as in the bulk phase, the very small differences in $\Delta$ between the strained phases means that it cannot be the only controlling feature; instead there is a complex interplay between the various structural descriptors which produces  the observed ground state.
The fact that Ca$_2$Fe$_2$O$_5$ has a smaller pseudo-cubic lattice parameter than Sr$_2$Fe$_2$O$_5$ means that the parallel phase is experimentally accessible with much more modest strain.
Experimental results on thin films of Ca$_2$Fe$_2$O$_5$ show that the vacancies order perpendicularly when grown on LaSrAlO$_4$ (LSAO, $a_{pc}$ = 3.75 \AA) and LaAlO$_3$ ($a_{pc}$ = 3.79 \AA), but order parallel on LSAT ($a_{pc}$ = 3.87 \AA) and SrTiO$_3$ ($a_{pc}$ = 3.91 \AA), again in agreement with these theoretical results.\cite{Rossell/Lebedev_etal:2004,Inoue_etal:2010}
Interestingly, the stabilization of the perpendicular phase under compressive strain and parallel phase under tensile strain does not hold across all brownmillerite oxide chemistries; in the cobaltates, for example, the trend is reversed.

\begin{figure*}
\centering
\includegraphics[width=1.8\columnwidth,clip]{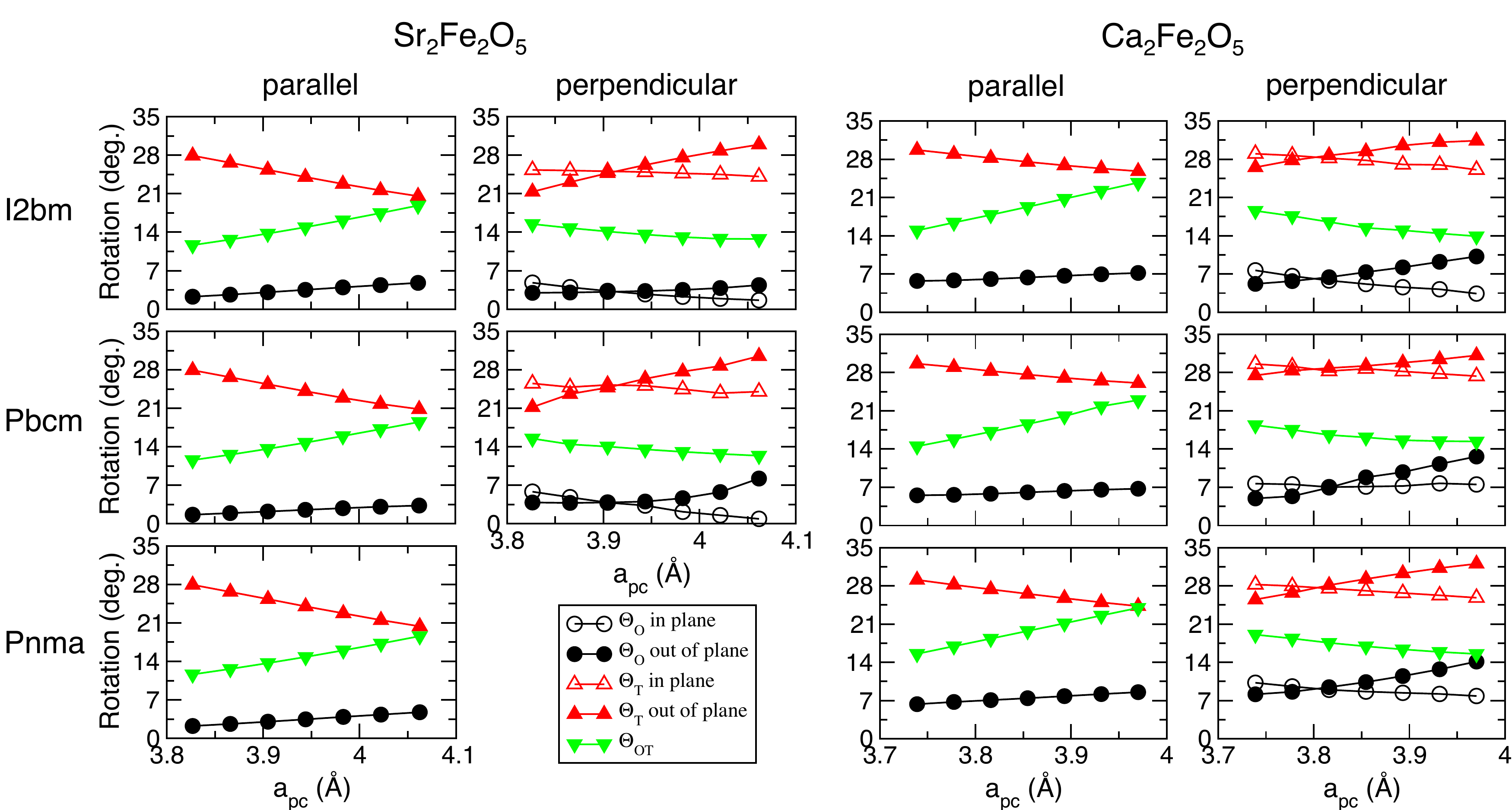}\vspace{-1\baselineskip}
\caption{The effect of strain on the tilt angles (as defined in Figure \ref{fig:angles_bulk}) of the different Sr$_2$Fe$_2$O$_5$ (left) and Ca$_2$Fe$_2$O$_5$ (right) structures with both parallel and perpendicular vacancy orientation. 
}
\label{fig:strainstructure}
\end{figure*}

We next look at the structural evolution of these compounds under strain.
For the structures with the parallel orientation, there are only the three bond angles defined previously to consider (Figure \ref{fig:angles_bulk}).
Under increasing tensile strain, the FeO$_4$ tetrahedral chains become less distorted (decreasing Fe-O-Fe bond angle), while the octahedral network and connections between the tetrahedral and octahedral layers, becomes more distorted (increasing Fe-O-Fe bond angle).
It should come as no surprise that this trend holds across all three types of ordered phases, as they are all very structural similar with the exception of the relative orientation of tetrahedral chains (Figure \ref{fig:strainstructure}).
Additionally, the chemistry of the $A$-site, although influencing the magnitude of the tilts, does not affect this trend.

When the oxygen-deficient layers reorient to become perpendicular to the substrate, there are now two more angles to consider.
In the previous geometries, only one angle of the tetrahedra and octahedra had to be considered owing to the fact that the equatorial oxygen atoms of each were in the plane of epitaxial strain.
Now these same atoms are oriented such that they are affected by both the plane of epitaxial strain and the strain-induced changes to the out-of-plane lattice parameter.
In order to get a full understanding of the structural distortions in this case, we separate the tetrahedral and octahedral angles into in-plane and out-of-plane components.
In this perpendicular orientation, increasing tensile strain increases the out-of-plane component of the octahedral and tetrahedral angles, while decreasing the in-plane component, as well as the angle between the layers.

\begin{figure*}
\centering
\includegraphics[width=1.9\columnwidth,clip]{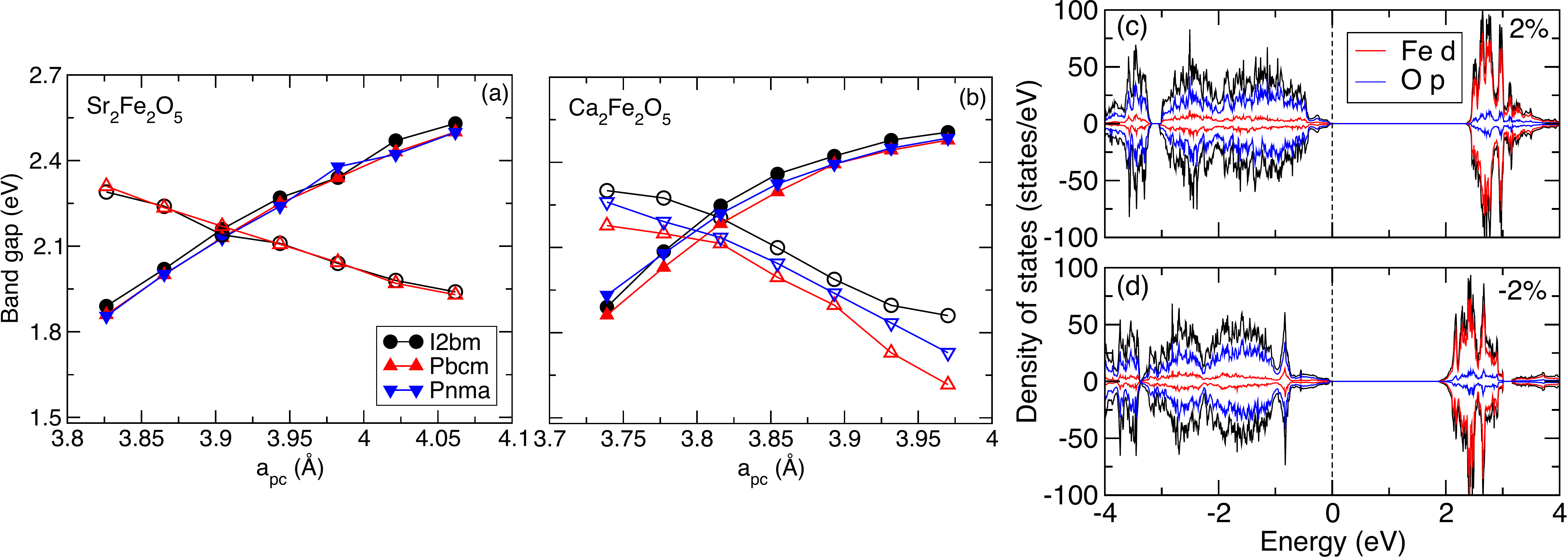}
\caption{The band gap of (a) Sr$_2$Fe$_2$O$_5$ and (b) Ca$_2$Fe$_2$O$_5$ are strongly influenced by epitaxial strain; the electronic gap for thin films structures with vacancies ordered parallel to the substrate (filled symbols) increases as tensile strain increases, but decreases in the perpendicular orientation (empty symbols). This electronic structure response occurs owing to the manner in which strain affects the angle between the tetrahedral and octahedral layers. The $Pbcm$ structure of Sr$_2$Fe$_2$O$_5$ with vacancies parallel to the substrate, for example, has a $\sim$ 0.3\,eV larger band gap under (c) 2\% tensile strain when compared to (d) 2\% compressive strain as seen by the change in the atom-resolved densities of states.
}
\label{fig:straindos}
\end{figure*}

Finally, the band gap of both materials is strongly influenced by strain, ranging from 1.8\,eV to 2.6\,eV for Sr$_2$Fe$_2$O$_5$ (Figure \ref{fig:straindos}a) and 1.6\,eV to 2.5\,eV for Ca$_2$Fe$_2$O$_5$ (Figure \ref{fig:straindos}b).
In both compounds, increasing tensile strain results in an increase of the band gap for the parallel orientation of vacancies, but interestingly, a decrease in the perpendicular orientation.
This is due to how an increase in the lattice parameters influences the connectivity of the tetrahedrally and octahedral coordinated iron atoms (\textit{i.e.}, $\Theta_{OT}$).
For the films with the parallel vacancy orientation, $\Theta_{OT}$ deviates further away from 180$^{\circ}$ as tensile strain increases; this results in decreased overlap of the Fe $d$-orbitals and O $p$-orbitals, giving a higher band gap.
Figures \ref{fig:straindos}c and \ref{fig:straindos}d show, as an example, the difference in the band gap of parallel-oriented Sr$_2$Fe$_2$O$_5$ at 2\% (2.43\,eV) and -2\% (2.00\,eV).
The opposite effect occurs in the perpendicular case, where a decrease in $\Theta_{OT}$ provides better overlap and thus a smaller band gap through increased bandwidth.

\subsection{Comparison to Perovskite Oxides}

\begin{figure}
\centering
\includegraphics[width=0.95\columnwidth,clip]{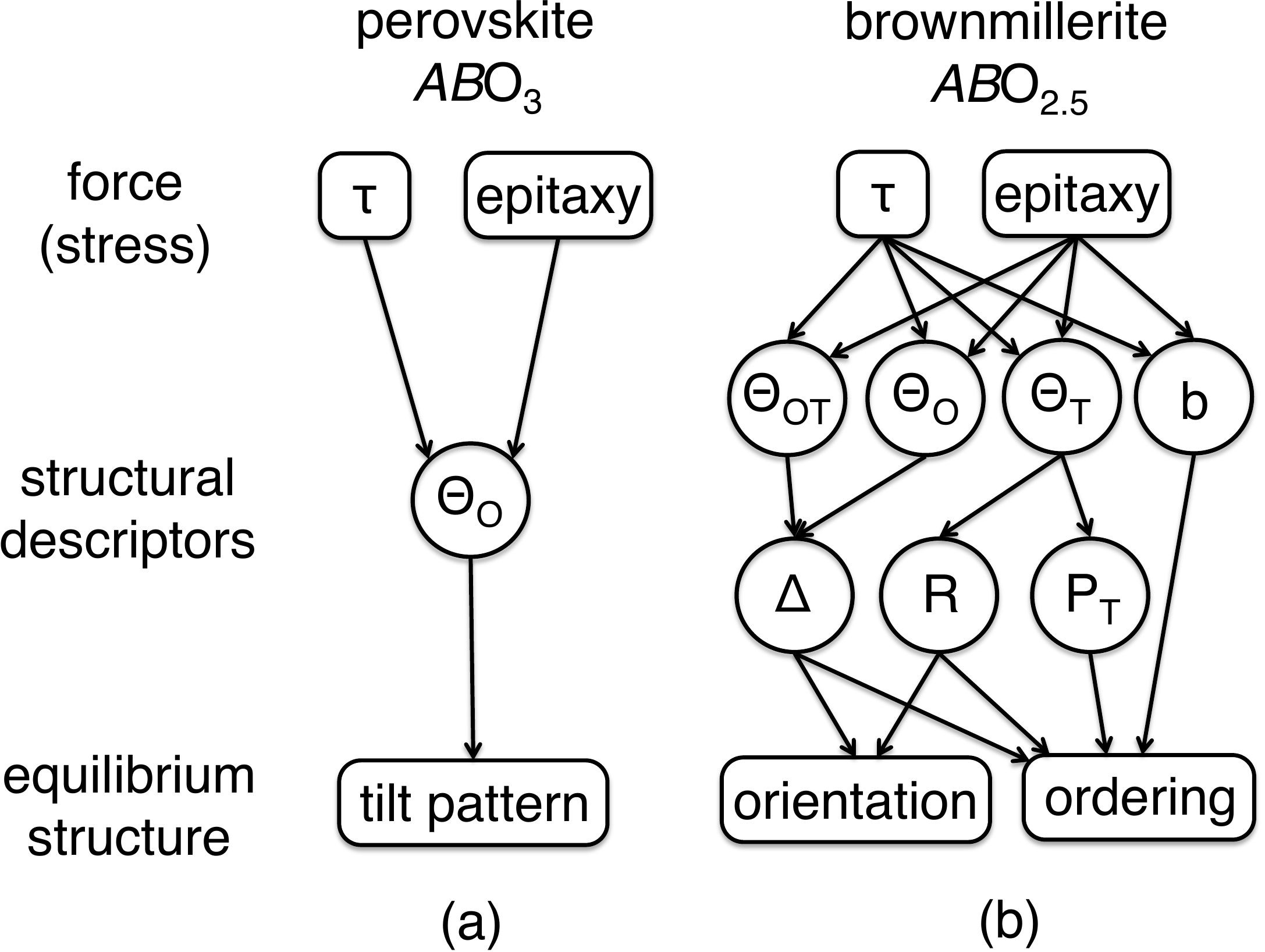}
\caption{The effect of different stress stimuli (ionic size or chemical pressure [captured by $\tau$] and epitaxy) on various structure descriptors,  which combine to produce the equilibrium crystal structure in (a) perovskite and (b) brownmillerite oxides.
}
\label{fig:flowchart}
\end{figure}

The fact that $B$O$_6$ octahedra in perovskite oxides form a flexible corner-connected network (Figure \ref{fig:bulk}a) allows them to easily rotate in space about the different crystallographic axes.
The size of these rotations directly affect the magnitude of the $B$--O--$B$ bond angles, which in turn impacts many electronic and magnetic properties.
Although there are 15 distinct ways in which the octahedra can cooperatively rotate while retaining connectivity (as identified by Glazer),\cite{Glazer:1972} the vast majority exhibit either an orthorhombic or rhombohedral tilt pattern (given by $a^-a^-c^+$ or $a^-a^-a^-$ in Glazer notation, respectively).
A high degree of control over the electronic structure can be achieved by using epitaxial strain (or chemical substation, as captured by $\tau$) to control these rotations, owing primarily to the strong coupling between the lattice and electronic degrees of freedom in perovskites (Figure \ref{fig:flowchart}a);\cite{Vailionis_etal:2011,Rondinelli/Spaldin:2011,Rondinelli/May/Freeland:2012,Spaldin/Schlom_etal:2013} as mentioned previously, buckling of the $B$--O--$B$ bond away from a linear 180$^{\circ}$ configuration (\textit{i.e.}, increasing the magnitude of the octahedral rotations) decreases the overlap between the O $p$ and metal $B$ $d$ orbitals, thereby increasing the band gap in insulating compounds or inducing bandwidth-driven metal-insulator transitions.\cite{Eng/Woodward:2003,Amat_etal:2014,Filip_etal:2014,Aschauer/Spaldin:2014,Li_etal:2015,Imada/Fujimori/Tokura:1998}

As we have shown, the distinct alternating tetrahedral and octahedral layers in brownmillerites allows for many more structural degrees of freedom than perovskites.
Application of the same forces (chemical pressure and epitaxial strain) now affects the interlayer separation of tetrahedra (\emph{i.e.}, $b$ lattice parameter), as well as the rotations (connectivity) of the different polyhedra (given by $\Theta_{O}$, $\Theta_{T}$, and $\Theta_{OT}$); each of these in turn influences the octahedral distortions ($\Delta$), the intralayer dipole separation ($R$), and the magnitude of the local tetrahedral dipoles ($P_T$).
The balance of these different factors, which is shown schematically in Figure \ref{fig:flowchart}b, 
is then what governs the equilibrium structure---the tetrahedral chain ordering and, for the case of thin films, the 
orientation of the vacancies relative to the substrate.

A feature implicitly indicated in Figure \ref{fig:flowchart} is that the electronic properties of these oxides are 
also a structural consequence; hence, control over the structural descriptors makes it possible to 
control the electronic response. 
Indeed, the band gaps of the brownmillerites are highly sensitive to rotations of the polyhedra; like the perovskites, the gap opens as the rotations increase owing to a change in orbital overlap.
Unlike perovskites, however, it is not solely a single $B$--O--$B$ bond angle controlling the electronic structure, which is due to the breaking of the structural topology by the oxygen-vacant (tetrahedral) layers.
For the brownmillerite oxides, it is the out-of-plane angle between the $B$O$_4$ tetrahedra and $B$O$_6$ octahedra, given by $\Theta_{OT}$ (or, alternatively, $B_\mathrm{tet}$--O--$B_\mathrm{oct}$), that controls the gap and band edge character.
Furthermore, strain affects the band gap of parallel or perpendicularly oriented brownmillerites in completely opposite (asymmetric) ways owing to the response of $\Theta_{OT}$ to the biaxial strain state.
Although strain has been used to control functional properties in perovskites, it appears to induce even more interesting responses in brownmillerites owing to the additional structural degrees of freedom present.

\section{Conclusion}

Using first principles density functional theory calculations, we investigated the brownmillerite compounds Sr$_2$Fe$_2$O$_5$ and Ca$_2$Fe$_2$O$_5$.
These anion deficient materials offer many more degrees of freedom than the related perovskite structures, increasing the structural complexity greatly.
The symmetry of these structures is determined by the relative ordering of left- and right-handed tetrahedral chains; which ordering scheme preferred, however, is determined by the interplay of several structural factors.
Although previous studies have attempted to characterize the various members of this family, 
here we clearly demonstrate how the tolerance factors, polarizations generated by rotations of FeO$_4$ tetrahedra, distortions of FeO$_6$ octahedra, and intralayer chain separation interact to produce the observed equilibrium phases.

Furthermore, we show the crystal and electronic structure of these phases are highly tunable using epitaxial strain.
We find that under compressive strain, the preferred ordering of vacancies (tetrahedral FeO$_4$ chains) 
changes from parallel to perpendicular when the separation between the change becomes better maximized.
The rotation angles are also strongly influenced, resulting in a increase (or decrease) in the band gap for the parallel (or perpendicular) orientation.
While this study is by no means comprehensive in scope, we hope that the framework presented here encourages the consideration of additional structural metrics (especially intralayer separation) in future studies and helps provide greater insight into this complex family of materials.
We can now harness this knowledge and understanding of the atomistic mechanisms behind the coupling of the crystal and electronic structure in brownmillerites to design electronic function and realize new materials platforms for thin film devices.

\section{Acknowledgments}

We wish to thank members of the Materials Theory and Design Group, as well as D. Fong, for useful discussions.
J.Y.\ and J.M.R.\ were supported by the U.S.\ DOE, Office of Basic Energy Sciences (BES), DE-AC02-06CH11357.
DFT calculations were performed on hardware supported by Drexel's University Research Computing Facility, and the CARBON cluster at the Center for Nanoscale Materials (Argonne National Laboratory, also supported by DOE-BES DE-AC02-06CH11357) under allocation CNM39812.

\bibliography{brownmillerites}

\begin{thebibliography}{55}%
\makeatletter
\providecommand \@ifxundefined [1]{%
 \@ifx{#1\undefined}
}%
\providecommand \@ifnum [1]{%
 \ifnum #1\expandafter \@firstoftwo
 \else \expandafter \@secondoftwo
 \fi
}%
\providecommand \@ifx [1]{%
 \ifx #1\expandafter \@firstoftwo
 \else \expandafter \@secondoftwo
 \fi
}%
\providecommand \natexlab [1]{#1}%
\providecommand \enquote  [1]{``#1''}%
\providecommand \bibnamefont  [1]{#1}%
\providecommand \bibfnamefont [1]{#1}%
\providecommand \citenamefont [1]{#1}%
\providecommand \href@noop [0]{\@secondoftwo}%
\providecommand \href [0]{\begingroup \@sanitize@url \@href}%
\providecommand \@href[1]{\@@startlink{#1}\@@href}%
\providecommand \@@href[1]{\endgroup#1\@@endlink}%
\providecommand \@sanitize@url [0]{\catcode `\\12\catcode `\$12\catcode
  `\&12\catcode `\#12\catcode `\^12\catcode `\_12\catcode `\%12\relax}%
\providecommand \@@startlink[1]{}%
\providecommand \@@endlink[0]{}%
\providecommand \url  [0]{\begingroup\@sanitize@url \@url }%
\providecommand \@url [1]{\endgroup\@href {#1}{\urlprefix }}%
\providecommand \urlprefix  [0]{URL }%
\providecommand \Eprint [0]{\href }%
\providecommand \doibase [0]{http://dx.doi.org/}%
\providecommand \selectlanguage [0]{\@gobble}%
\providecommand \bibinfo  [0]{\@secondoftwo}%
\providecommand \bibfield  [0]{\@secondoftwo}%
\providecommand \translation [1]{[#1]}%
\providecommand \BibitemOpen [0]{}%
\providecommand \bibitemStop [0]{}%
\providecommand \bibitemNoStop [0]{.\EOS\space}%
\providecommand \EOS [0]{\spacefactor3000\relax}%
\providecommand \BibitemShut  [1]{\csname bibitem#1\endcsname}%
\let\auto@bib@innerbib\@empty
\bibitem [{\citenamefont {Kendall}\ \emph {et~al.}(1995)\citenamefont
  {Kendall}, \citenamefont {Navas}, \citenamefont {Thomas},\ and\ \citenamefont
  {zur Loye}}]{Kendall_etal:1995}%
  \BibitemOpen
  \bibfield  {author} {\bibinfo {author} {\bibfnamefont {K.~R.}\ \bibnamefont
  {Kendall}}, \bibinfo {author} {\bibfnamefont {C.}~\bibnamefont {Navas}},
  \bibinfo {author} {\bibfnamefont {J.~K.}\ \bibnamefont {Thomas}}, \ and\
  \bibinfo {author} {\bibfnamefont {H.-C.}\ \bibnamefont {zur Loye}},\
  }\href@noop {} {\bibfield  {journal} {\bibinfo  {journal} {Solid State
  Ionics}\ }\textbf {\bibinfo {volume} {82}},\ \bibinfo {pages} {215} (\bibinfo
  {year} {1995})}\BibitemShut {NoStop}%
\bibitem [{\citenamefont {Boivin}\ and\ \citenamefont
  {Mairesse}(1998)}]{Boivin/Mairesse:1998}%
  \BibitemOpen
  \bibfield  {author} {\bibinfo {author} {\bibfnamefont {J.~C.}\ \bibnamefont
  {Boivin}}\ and\ \bibinfo {author} {\bibfnamefont {G.}~\bibnamefont
  {Mairesse}},\ }\href@noop {} {\bibfield  {journal} {\bibinfo  {journal}
  {Chem. Mater.}\ }\textbf {\bibinfo {volume} {10}},\ \bibinfo {pages} {2870}
  (\bibinfo {year} {1998})}\BibitemShut {NoStop}%
\bibitem [{\citenamefont {Rolle}\ \emph {et~al.}(2005)\citenamefont {Rolle},
  \citenamefont {Vannier}, \citenamefont {Giridharan},\ and\ \citenamefont
  {Abraham}}]{Rolle_etal:2005}%
  \BibitemOpen
  \bibfield  {author} {\bibinfo {author} {\bibfnamefont {A.}~\bibnamefont
  {Rolle}}, \bibinfo {author} {\bibfnamefont {R.~N.}\ \bibnamefont {Vannier}},
  \bibinfo {author} {\bibfnamefont {N.~V.}\ \bibnamefont {Giridharan}}, \ and\
  \bibinfo {author} {\bibfnamefont {F.}~\bibnamefont {Abraham}},\ }\href@noop
  {} {\bibfield  {journal} {\bibinfo  {journal} {Solid State Ionics}\ }\textbf
  {\bibinfo {volume} {176}},\ \bibinfo {pages} {2095} (\bibinfo {year}
  {2005})}\BibitemShut {NoStop}%
\bibitem [{\citenamefont {Shaula}\ \emph {et~al.}(2006)\citenamefont {Shaula},
  \citenamefont {Pivak}, \citenamefont {Waerenborgh}, \citenamefont
  {Gaczy\~{n}ski}, \citenamefont {Yaremchenko},\ and\ \citenamefont
  {Karton}}]{Shaula_etal:2006}%
  \BibitemOpen
  \bibfield  {author} {\bibinfo {author} {\bibfnamefont {A.~L.}\ \bibnamefont
  {Shaula}}, \bibinfo {author} {\bibfnamefont {Y.~V.}\ \bibnamefont {Pivak}},
  \bibinfo {author} {\bibfnamefont {J.~C.}\ \bibnamefont {Waerenborgh}},
  \bibinfo {author} {\bibfnamefont {P.}~\bibnamefont {Gaczy\~{n}ski}}, \bibinfo
  {author} {\bibfnamefont {A.~A.}\ \bibnamefont {Yaremchenko}}, \ and\ \bibinfo
  {author} {\bibfnamefont {V.~V.}\ \bibnamefont {Karton}},\ }\href@noop {}
  {\bibfield  {journal} {\bibinfo  {journal} {Solid State Ionics}\ }\textbf
  {\bibinfo {volume} {177}},\ \bibinfo {pages} {2923} (\bibinfo {year}
  {2006})}\BibitemShut {NoStop}%
\bibitem [{\citenamefont {Orera}\ and\ \citenamefont
  {Slater}(2010)}]{Orera/Slater:2010}%
  \BibitemOpen
  \bibfield  {author} {\bibinfo {author} {\bibfnamefont {A.}~\bibnamefont
  {Orera}}\ and\ \bibinfo {author} {\bibfnamefont {P.~R.}\ \bibnamefont
  {Slater}},\ }\href@noop {} {\bibfield  {journal} {\bibinfo  {journal} {Chem.
  Mater.}\ }\textbf {\bibinfo {volume} {22}},\ \bibinfo {pages} {675} (\bibinfo
  {year} {2010})}\BibitemShut {NoStop}%
\bibitem [{\citenamefont {Antipov}\ \emph {et~al.}(2004)\citenamefont
  {Antipov}, \citenamefont {Abakumov}, \citenamefont {Alekseeva}, \citenamefont
  {Rozova}, \citenamefont {Hadermann}, \citenamefont {Lebedev},\ and\
  \citenamefont {Van~Tendeloo}}]{Antipov_etal:2004}%
  \BibitemOpen
  \bibfield  {author} {\bibinfo {author} {\bibfnamefont {E.~V.}\ \bibnamefont
  {Antipov}}, \bibinfo {author} {\bibfnamefont {A.~M.}\ \bibnamefont
  {Abakumov}}, \bibinfo {author} {\bibfnamefont {A.~M.}\ \bibnamefont
  {Alekseeva}}, \bibinfo {author} {\bibfnamefont {M.~G.}\ \bibnamefont
  {Rozova}}, \bibinfo {author} {\bibfnamefont {J.}~\bibnamefont {Hadermann}},
  \bibinfo {author} {\bibfnamefont {O.~I.}\ \bibnamefont {Lebedev}}, \ and\
  \bibinfo {author} {\bibfnamefont {G.}~\bibnamefont {Van~Tendeloo}},\
  }\href@noop {} {\bibfield  {journal} {\bibinfo  {journal} {Phys. Status
  Solidi A}\ }\textbf {\bibinfo {volume} {201}},\ \bibinfo {pages} {1403}
  (\bibinfo {year} {2004})}\BibitemShut {NoStop}%
\bibitem [{\citenamefont {Sullivan}\ and\ \citenamefont
  {Greaves}(2012)}]{Sullivan/Greaves:2012}%
  \BibitemOpen
  \bibfield  {author} {\bibinfo {author} {\bibfnamefont {E.}~\bibnamefont
  {Sullivan}}\ and\ \bibinfo {author} {\bibfnamefont {C.}~\bibnamefont
  {Greaves}},\ }\href@noop {} {\bibfield  {journal} {\bibinfo  {journal}
  {Mater. Res. Bull.}\ }\textbf {\bibinfo {volume} {47}},\ \bibinfo {pages}
  {2541} (\bibinfo {year} {2012})}\BibitemShut {NoStop}%
\bibitem [{\citenamefont {Tarasova}\ \emph {et~al.}(2013)\citenamefont
  {Tarasova}, \citenamefont {Filinkova},\ and\ \citenamefont
  {Animitsa}}]{Tarasova_etal:2013}%
  \BibitemOpen
  \bibfield  {author} {\bibinfo {author} {\bibfnamefont {N.~A.}\ \bibnamefont
  {Tarasova}}, \bibinfo {author} {\bibfnamefont {Y.~V.}\ \bibnamefont
  {Filinkova}}, \ and\ \bibinfo {author} {\bibfnamefont {I.~E.}\ \bibnamefont
  {Animitsa}},\ }\href@noop {} {\bibfield  {journal} {\bibinfo  {journal}
  {Russ. J. Electrochem.}\ }\textbf {\bibinfo {volume} {49}},\ \bibinfo {pages}
  {45} (\bibinfo {year} {2013})}\BibitemShut {NoStop}%
\bibitem [{\citenamefont {Battle}\ \emph {et~al.}(1988)\citenamefont {Battle},
  \citenamefont {Gibb},\ and\ \citenamefont {Lightfoot}}]{Battle_etal:1988}%
  \BibitemOpen
  \bibfield  {author} {\bibinfo {author} {\bibfnamefont {P.~D.}\ \bibnamefont
  {Battle}}, \bibinfo {author} {\bibfnamefont {T.~C.}\ \bibnamefont {Gibb}}, \
  and\ \bibinfo {author} {\bibfnamefont {P.}~\bibnamefont {Lightfoot}},\
  }\href@noop {} {\bibfield  {journal} {\bibinfo  {journal} {J. Solid State
  Chem.}\ }\textbf {\bibinfo {volume} {76}},\ \bibinfo {pages} {334} (\bibinfo
  {year} {1988})}\BibitemShut {NoStop}%
\bibitem [{\citenamefont {Abakumov}\ \emph {et~al.}(2003)\citenamefont
  {Abakumov}, \citenamefont {Alekseeva}, \citenamefont {Rozova}, \citenamefont
  {Antipov}, \citenamefont {Lebedev},\ and\ \citenamefont
  {Van~Tendeloo}}]{Abakumov_etal:2003}%
  \BibitemOpen
  \bibfield  {author} {\bibinfo {author} {\bibfnamefont {A.~M.}\ \bibnamefont
  {Abakumov}}, \bibinfo {author} {\bibfnamefont {A.~M.}\ \bibnamefont
  {Alekseeva}}, \bibinfo {author} {\bibfnamefont {M.~G.}\ \bibnamefont
  {Rozova}}, \bibinfo {author} {\bibfnamefont {E.~V.}\ \bibnamefont {Antipov}},
  \bibinfo {author} {\bibfnamefont {O.~I.}\ \bibnamefont {Lebedev}}, \ and\
  \bibinfo {author} {\bibfnamefont {G.~V.}\ \bibnamefont {Van~Tendeloo}},\
  }\href@noop {} {\bibfield  {journal} {\bibinfo  {journal} {J. Solid State
  Chem.}\ }\textbf {\bibinfo {volume} {174}},\ \bibinfo {pages} {319} (\bibinfo
  {year} {2003})}\BibitemShut {NoStop}%
\bibitem [{\citenamefont {Lazic}\ \emph {et~al.}(2008)\citenamefont {Lazic},
  \citenamefont {Kr\"{u}her}, \citenamefont {Kahlenberg}, \citenamefont
  {Konzett},\ and\ \citenamefont {Kaindl}}]{Lazic_etal:2008}%
  \BibitemOpen
  \bibfield  {author} {\bibinfo {author} {\bibfnamefont {B.}~\bibnamefont
  {Lazic}}, \bibinfo {author} {\bibfnamefont {H.}~\bibnamefont {Kr\"{u}her}},
  \bibinfo {author} {\bibfnamefont {V.}~\bibnamefont {Kahlenberg}}, \bibinfo
  {author} {\bibfnamefont {J.}~\bibnamefont {Konzett}}, \ and\ \bibinfo
  {author} {\bibfnamefont {R.}~\bibnamefont {Kaindl}},\ }\href@noop {}
  {\bibfield  {journal} {\bibinfo  {journal} {Acta Cryst.}\ }\textbf {\bibinfo
  {volume} {B64}},\ \bibinfo {pages} {417} (\bibinfo {year}
  {2008})}\BibitemShut {NoStop}%
\bibitem [{\citenamefont {Kr\"{u}her}\ and\ \citenamefont
  {Kahlenberg}(2005)}]{Kruger/Kahlenberg:2005}%
  \BibitemOpen
  \bibfield  {author} {\bibinfo {author} {\bibfnamefont {H.}~\bibnamefont
  {Kr\"{u}her}}\ and\ \bibinfo {author} {\bibfnamefont {V.}~\bibnamefont
  {Kahlenberg}},\ }\href@noop {} {\bibfield  {journal} {\bibinfo  {journal}
  {Acta Cryst.}\ }\textbf {\bibinfo {volume} {B61}},\ \bibinfo {pages} {656}
  (\bibinfo {year} {2005})}\BibitemShut {NoStop}%
\bibitem [{\citenamefont {Klenov}\ \emph {et~al.}(2003)\citenamefont {Klenov},
  \citenamefont {Donner}, \citenamefont {Foran},\ and\ \citenamefont
  {Stemmer}}]{Klenov_etal:2003}%
  \BibitemOpen
  \bibfield  {author} {\bibinfo {author} {\bibfnamefont {D.~O.}\ \bibnamefont
  {Klenov}}, \bibinfo {author} {\bibfnamefont {W.}~\bibnamefont {Donner}},
  \bibinfo {author} {\bibfnamefont {B.}~\bibnamefont {Foran}}, \ and\ \bibinfo
  {author} {\bibfnamefont {S.}~\bibnamefont {Stemmer}},\ }\href@noop {}
  {\bibfield  {journal} {\bibinfo  {journal} {Appl. Phys. Lett.}\ }\textbf
  {\bibinfo {volume} {82}},\ \bibinfo {pages} {3427} (\bibinfo {year}
  {2003})}\BibitemShut {NoStop}%
\bibitem [{\citenamefont {Gazquez}\ \emph {et~al.}(2013)\citenamefont
  {Gazquez}, \citenamefont {Bose}, \citenamefont {Sharma}, \citenamefont
  {Torija}, \citenamefont {Pennycook}, \citenamefont {Leighton},\ and\
  \citenamefont {Varela}}]{Gazquez_etal:2013}%
  \BibitemOpen
  \bibfield  {author} {\bibinfo {author} {\bibfnamefont {J.}~\bibnamefont
  {Gazquez}}, \bibinfo {author} {\bibfnamefont {S.}~\bibnamefont {Bose}},
  \bibinfo {author} {\bibfnamefont {M.}~\bibnamefont {Sharma}}, \bibinfo
  {author} {\bibfnamefont {M.~A.}\ \bibnamefont {Torija}}, \bibinfo {author}
  {\bibfnamefont {S.~J.}\ \bibnamefont {Pennycook}}, \bibinfo {author}
  {\bibfnamefont {C.}~\bibnamefont {Leighton}}, \ and\ \bibinfo {author}
  {\bibfnamefont {M.}~\bibnamefont {Varela}},\ }\href@noop {} {\bibfield
  {journal} {\bibinfo  {journal} {APL Mater.}\ }\textbf {\bibinfo {volume}
  {1}},\ \bibinfo {pages} {012105} (\bibinfo {year} {2013})}\BibitemShut
  {NoStop}%
\bibitem [{\citenamefont {Inoue}\ \emph {et~al.}(2010)\citenamefont {Inoue},
  \citenamefont {Kawai}, \citenamefont {Ichikawa}, \citenamefont {Kageyama},
  \citenamefont {Paulus},\ and\ \citenamefont {Shimakawa}}]{Inoue_etal:2010}%
  \BibitemOpen
  \bibfield  {author} {\bibinfo {author} {\bibfnamefont {S.}~\bibnamefont
  {Inoue}}, \bibinfo {author} {\bibfnamefont {M.}~\bibnamefont {Kawai}},
  \bibinfo {author} {\bibfnamefont {N.}~\bibnamefont {Ichikawa}}, \bibinfo
  {author} {\bibfnamefont {H.}~\bibnamefont {Kageyama}}, \bibinfo {author}
  {\bibfnamefont {W.}~\bibnamefont {Paulus}}, \ and\ \bibinfo {author}
  {\bibfnamefont {Y.}~\bibnamefont {Shimakawa}},\ }\href@noop {} {\bibfield
  {journal} {\bibinfo  {journal} {Nature Chem.}\ }\textbf {\bibinfo {volume}
  {2}},\ \bibinfo {pages} {213} (\bibinfo {year} {2010})}\BibitemShut {NoStop}%
\bibitem [{\citenamefont {Schmidt}\ and\ \citenamefont
  {Campbell}(2001)}]{Schmidt/Campbell:2001}%
  \BibitemOpen
  \bibfield  {author} {\bibinfo {author} {\bibfnamefont {M.}~\bibnamefont
  {Schmidt}}\ and\ \bibinfo {author} {\bibfnamefont {S.~J.}\ \bibnamefont
  {Campbell}},\ }\href@noop {} {\bibfield  {journal} {\bibinfo  {journal} {J.
  Solid State Chem.}\ }\textbf {\bibinfo {volume} {156}},\ \bibinfo {pages}
  {292} (\bibinfo {year} {2001})}\BibitemShut {NoStop}%
\bibitem [{\citenamefont {Kr\"{u}ger}\ \emph {et~al.}(2009)\citenamefont
  {Kr\"{u}ger}, \citenamefont {Kahlenberg}, \citenamefont
  {Pet\v{r}\'{i}\v{c}ek}, \citenamefont {Phillipp},\ and\ \citenamefont
  {Wertl}}]{Kruger_2009}%
  \BibitemOpen
  \bibfield  {author} {\bibinfo {author} {\bibfnamefont {H.}~\bibnamefont
  {Kr\"{u}ger}}, \bibinfo {author} {\bibfnamefont {V.}~\bibnamefont
  {Kahlenberg}}, \bibinfo {author} {\bibfnamefont {V.}~\bibnamefont
  {Pet\v{r}\'{i}\v{c}ek}}, \bibinfo {author} {\bibfnamefont {F.}~\bibnamefont
  {Phillipp}}, \ and\ \bibinfo {author} {\bibfnamefont {W.}~\bibnamefont
  {Wertl}},\ }\href@noop {} {\bibfield  {journal} {\bibinfo  {journal} {J.
  Solid State Chem.}\ }\textbf {\bibinfo {volume} {182}},\ \bibinfo {pages}
  {1515} (\bibinfo {year} {2009})}\BibitemShut {NoStop}%
\bibitem [{\citenamefont {Auckett}\ \emph {et~al.}(2012)\citenamefont
  {Auckett}, \citenamefont {Studer}, \citenamefont {Sharma},\ and\
  \citenamefont {Ling}}]{Auckett_etal:2012}%
  \BibitemOpen
  \bibfield  {author} {\bibinfo {author} {\bibfnamefont {J.~E.}\ \bibnamefont
  {Auckett}}, \bibinfo {author} {\bibfnamefont {A.~J.}\ \bibnamefont {Studer}},
  \bibinfo {author} {\bibfnamefont {N.}~\bibnamefont {Sharma}}, \ and\ \bibinfo
  {author} {\bibfnamefont {C.~D.}\ \bibnamefont {Ling}},\ }\href@noop {}
  {\bibfield  {journal} {\bibinfo  {journal} {Solid State Ionics}\ }\textbf
  {\bibinfo {volume} {225}},\ \bibinfo {pages} {432} (\bibinfo {year}
  {2012})}\BibitemShut {NoStop}%
\bibitem [{\citenamefont {Auckett}\ \emph {et~al.}(2013)\citenamefont
  {Auckett}, \citenamefont {Studer}, \citenamefont {Pellegrini}, \citenamefont
  {Ollivier}, \citenamefont {Johnson}, \citenamefont {Schober}, \citenamefont
  {Miiller},\ and\ \citenamefont {Ling}}]{Auckett_etal:2013}%
  \BibitemOpen
  \bibfield  {author} {\bibinfo {author} {\bibfnamefont {J.~E.}\ \bibnamefont
  {Auckett}}, \bibinfo {author} {\bibfnamefont {A.~J.}\ \bibnamefont {Studer}},
  \bibinfo {author} {\bibfnamefont {E.}~\bibnamefont {Pellegrini}}, \bibinfo
  {author} {\bibfnamefont {J.}~\bibnamefont {Ollivier}}, \bibinfo {author}
  {\bibfnamefont {M.~R.}\ \bibnamefont {Johnson}}, \bibinfo {author}
  {\bibfnamefont {H.}~\bibnamefont {Schober}}, \bibinfo {author} {\bibfnamefont
  {W.}~\bibnamefont {Miiller}}, \ and\ \bibinfo {author} {\bibfnamefont
  {C.~D.}\ \bibnamefont {Ling}},\ }\href@noop {} {\bibfield  {journal}
  {\bibinfo  {journal} {Chem. Mater.}\ }\textbf {\bibinfo {volume} {25}},\
  \bibinfo {pages} {3080} (\bibinfo {year} {2013})}\BibitemShut {NoStop}%
\bibitem [{\citenamefont {Takeda}\ \emph {et~al.}(1968)\citenamefont {Takeda},
  \citenamefont {Yamaguchi}, \citenamefont {Tomiyoshi}, \citenamefont {Fukase},
  \citenamefont {Sugimoto},\ and\ \citenamefont
  {Yamamoto}}]{Takeda_etal_2:1968}%
  \BibitemOpen
  \bibfield  {author} {\bibinfo {author} {\bibfnamefont {T.}~\bibnamefont
  {Takeda}}, \bibinfo {author} {\bibfnamefont {Y.}~\bibnamefont {Yamaguchi}},
  \bibinfo {author} {\bibfnamefont {S.}~\bibnamefont {Tomiyoshi}}, \bibinfo
  {author} {\bibfnamefont {M.}~\bibnamefont {Fukase}}, \bibinfo {author}
  {\bibfnamefont {M.}~\bibnamefont {Sugimoto}}, \ and\ \bibinfo {author}
  {\bibfnamefont {H.}~\bibnamefont {Yamamoto}},\ }\href@noop {} {\bibfield
  {journal} {\bibinfo  {journal} {J. Phys. Soc. Jpn.}\ }\textbf {\bibinfo
  {volume} {24}},\ \bibinfo {pages} {446} (\bibinfo {year} {1968})}\BibitemShut
  {NoStop}%
\bibitem [{\citenamefont {Takeda}\ \emph {et~al.}(1969)\citenamefont {Takeda},
  \citenamefont {Yamaguchi}, \citenamefont {Watanabe}, \citenamefont
  {Tomiyoshi},\ and\ \citenamefont {Yamamoto}}]{Takeda_etal_1:1969}%
  \BibitemOpen
  \bibfield  {author} {\bibinfo {author} {\bibfnamefont {T.}~\bibnamefont
  {Takeda}}, \bibinfo {author} {\bibfnamefont {Y.}~\bibnamefont {Yamaguchi}},
  \bibinfo {author} {\bibfnamefont {J.}~\bibnamefont {Watanabe}}, \bibinfo
  {author} {\bibfnamefont {S.}~\bibnamefont {Tomiyoshi}}, \ and\ \bibinfo
  {author} {\bibfnamefont {H.}~\bibnamefont {Yamamoto}},\ }\href@noop {}
  {\bibfield  {journal} {\bibinfo  {journal} {J. Phys. Soc. Jpn.}\ }\textbf
  {\bibinfo {volume} {26}},\ \bibinfo {pages} {1320} (\bibinfo {year}
  {1969})}\BibitemShut {NoStop}%
\bibitem [{\citenamefont {Galakhov}\ \emph {et~al.}(2010)\citenamefont
  {Galakhov}, \citenamefont {Kurmaev}, \citenamefont {Kuepper}, \citenamefont
  {Neumann}, \citenamefont {McLeod}, \citenamefont {Moewes}, \citenamefont
  {Leonidov},\ and\ \citenamefont {Kozhevnikov}}]{Galakhov_etal:2010}%
  \BibitemOpen
  \bibfield  {author} {\bibinfo {author} {\bibfnamefont {V.~R.}\ \bibnamefont
  {Galakhov}}, \bibinfo {author} {\bibfnamefont {E.~Z.}\ \bibnamefont
  {Kurmaev}}, \bibinfo {author} {\bibfnamefont {K.}~\bibnamefont {Kuepper}},
  \bibinfo {author} {\bibfnamefont {M.}~\bibnamefont {Neumann}}, \bibinfo
  {author} {\bibfnamefont {J.~A.}\ \bibnamefont {McLeod}}, \bibinfo {author}
  {\bibfnamefont {A.}~\bibnamefont {Moewes}}, \bibinfo {author} {\bibfnamefont
  {I.~A.}\ \bibnamefont {Leonidov}}, \ and\ \bibinfo {author} {\bibfnamefont
  {V.~L.}\ \bibnamefont {Kozhevnikov}},\ }\href@noop {} {\bibfield  {journal}
  {\bibinfo  {journal} {J. Phys. Chem. C}\ }\textbf {\bibinfo {volume} {114}},\
  \bibinfo {pages} {5154} (\bibinfo {year} {2010})}\BibitemShut {NoStop}%
\bibitem [{\citenamefont {Hohenberg}\ and\ \citenamefont
  {Kohn}(1964)}]{Hohenberg/Kohn:1964}%
  \BibitemOpen
  \bibfield  {author} {\bibinfo {author} {\bibfnamefont {P.}~\bibnamefont
  {Hohenberg}}\ and\ \bibinfo {author} {\bibfnamefont {W.}~\bibnamefont
  {Kohn}},\ }\href@noop {} {\bibfield  {journal} {\bibinfo  {journal} {Phys.
  Rev.}\ }\textbf {\bibinfo {volume} {136}},\ \bibinfo {pages} {B864} (\bibinfo
  {year} {1964})}\BibitemShut {NoStop}%
\bibitem [{\citenamefont {Kresse}\ and\ \citenamefont
  {Hafner}(1993)}]{Kresse/Hafner:1993}%
  \BibitemOpen
  \bibfield  {author} {\bibinfo {author} {\bibfnamefont {G.}~\bibnamefont
  {Kresse}}\ and\ \bibinfo {author} {\bibfnamefont {J.}~\bibnamefont
  {Hafner}},\ }\href@noop {} {\bibfield  {journal} {\bibinfo  {journal} {Phys.
  Rev. B}\ }\textbf {\bibinfo {volume} {47}},\ \bibinfo {pages} {558} (\bibinfo
  {year} {1993})}\BibitemShut {NoStop}%
\bibitem [{\citenamefont {Kresse}\ and\ \citenamefont
  {Furthm\"uller}(1996)}]{Kresse/Hafner:1996}%
  \BibitemOpen
  \bibfield  {author} {\bibinfo {author} {\bibfnamefont {G.}~\bibnamefont
  {Kresse}}\ and\ \bibinfo {author} {\bibfnamefont {J.}~\bibnamefont
  {Furthm\"uller}},\ }\href@noop {} {\bibfield  {journal} {\bibinfo  {journal}
  {Comput. Mater. Sci.}\ }\textbf {\bibinfo {volume} {6}},\ \bibinfo {pages}
  {15} (\bibinfo {year} {1996})}\BibitemShut {NoStop}%
\bibitem [{\citenamefont {Bl\"ochl}(1994)}]{Blochl:1994}%
  \BibitemOpen
  \bibfield  {author} {\bibinfo {author} {\bibfnamefont {P.~E.}\ \bibnamefont
  {Bl\"ochl}},\ }\href@noop {} {\bibfield  {journal} {\bibinfo  {journal}
  {Phys. Rev. B}\ }\textbf {\bibinfo {volume} {50}},\ \bibinfo {pages} {17953}
  (\bibinfo {year} {1994})}\BibitemShut {NoStop}%
\bibitem [{\citenamefont {Perdew}\ \emph {et~al.}(2008)\citenamefont {Perdew},
  \citenamefont {Ruzsinszky}, \citenamefont {Csonka}, \citenamefont {Vydrov},
  \citenamefont {Scuseria}, \citenamefont {Constantin}, \citenamefont {Zhou},\
  and\ \citenamefont {Burke}}]{PBEsol:2008}%
  \BibitemOpen
  \bibfield  {author} {\bibinfo {author} {\bibfnamefont {J.~P.}\ \bibnamefont
  {Perdew}}, \bibinfo {author} {\bibfnamefont {A.}~\bibnamefont {Ruzsinszky}},
  \bibinfo {author} {\bibfnamefont {G.~I.}\ \bibnamefont {Csonka}}, \bibinfo
  {author} {\bibfnamefont {O.~A.}\ \bibnamefont {Vydrov}}, \bibinfo {author}
  {\bibfnamefont {G.~E.}\ \bibnamefont {Scuseria}}, \bibinfo {author}
  {\bibfnamefont {L.~A.}\ \bibnamefont {Constantin}}, \bibinfo {author}
  {\bibfnamefont {X.}~\bibnamefont {Zhou}}, \ and\ \bibinfo {author}
  {\bibfnamefont {K.}~\bibnamefont {Burke}},\ }\href@noop {} {\bibfield
  {journal} {\bibinfo  {journal} {Phys. Rev. Lett.}\ }\textbf {\bibinfo
  {volume} {100}},\ \bibinfo {pages} {136406} (\bibinfo {year}
  {2008})}\BibitemShut {NoStop}%
\bibitem [{\citenamefont {Monkhorst}\ and\ \citenamefont
  {Pack}(1976)}]{Monkhorst/Pack:1976}%
  \BibitemOpen
  \bibfield  {author} {\bibinfo {author} {\bibfnamefont {H.~J.}\ \bibnamefont
  {Monkhorst}}\ and\ \bibinfo {author} {\bibfnamefont {J.~D.}\ \bibnamefont
  {Pack}},\ }\href@noop {} {\bibfield  {journal} {\bibinfo  {journal} {Phys.
  Rev. B}\ }\textbf {\bibinfo {volume} {13}},\ \bibinfo {pages} {5188}
  (\bibinfo {year} {1976})}\BibitemShut {NoStop}%
\bibitem [{\citenamefont {Dudarev}\ \emph {et~al.}(1998)\citenamefont
  {Dudarev}, \citenamefont {Botton}, \citenamefont {Savrasov}, \citenamefont
  {Humphreys},\ and\ \citenamefont {Sutton}}]{Dudarev:1998}%
  \BibitemOpen
  \bibfield  {author} {\bibinfo {author} {\bibfnamefont {S.~L.}\ \bibnamefont
  {Dudarev}}, \bibinfo {author} {\bibfnamefont {G.~A.}\ \bibnamefont {Botton}},
  \bibinfo {author} {\bibfnamefont {S.~Y.}\ \bibnamefont {Savrasov}}, \bibinfo
  {author} {\bibfnamefont {C.~J.}\ \bibnamefont {Humphreys}}, \ and\ \bibinfo
  {author} {\bibfnamefont {A.~P.}\ \bibnamefont {Sutton}},\ }\href@noop {}
  {\bibfield  {journal} {\bibinfo  {journal} {Phys. Rev. B}\ }\textbf {\bibinfo
  {volume} {57}},\ \bibinfo {pages} {1505} (\bibinfo {year}
  {1998})}\BibitemShut {NoStop}%
\bibitem [{\citenamefont {Campbell}\ \emph {et~al.}(2006)\citenamefont
  {Campbell}, \citenamefont {Stokes}, \citenamefont {Tanner},\ and\
  \citenamefont {Hatch}}]{ISODISTORT}%
  \BibitemOpen
  \bibfield  {author} {\bibinfo {author} {\bibfnamefont {B.~J.}\ \bibnamefont
  {Campbell}}, \bibinfo {author} {\bibfnamefont {H.~T.}\ \bibnamefont
  {Stokes}}, \bibinfo {author} {\bibfnamefont {D.~E.}\ \bibnamefont {Tanner}},
  \ and\ \bibinfo {author} {\bibfnamefont {D.~M.}\ \bibnamefont {Hatch}},\
  }\href@noop {} {\bibfield  {journal} {\bibinfo  {journal} {J. Appl. Cryst.}\
  }\textbf {\bibinfo {volume} {39}},\ \bibinfo {pages} {607} (\bibinfo {year}
  {2006})}\BibitemShut {NoStop}%
\bibitem [{\citenamefont {Momma}\ and\ \citenamefont {Izumi}(2011)}]{VESTA}%
  \BibitemOpen
  \bibfield  {author} {\bibinfo {author} {\bibfnamefont {K.}~\bibnamefont
  {Momma}}\ and\ \bibinfo {author} {\bibfnamefont {F.}~\bibnamefont {Izumi}},\
  }\href@noop {} {\bibfield  {journal} {\bibinfo  {journal} {J. Appl. Cryst.}\
  }\textbf {\bibinfo {volume} {44}},\ \bibinfo {pages} {1272} (\bibinfo {year}
  {2011})}\BibitemShut {NoStop}%
\bibitem [{\citenamefont {Zayak}\ \emph {et~al.}(2006)\citenamefont {Zayak},
  \citenamefont {Huang}, \citenamefont {Neaton},\ and\ \citenamefont
  {Rabe}}]{Zayak_etal:2006}%
  \BibitemOpen
  \bibfield  {author} {\bibinfo {author} {\bibfnamefont {A.~T.}\ \bibnamefont
  {Zayak}}, \bibinfo {author} {\bibfnamefont {X.}~\bibnamefont {Huang}},
  \bibinfo {author} {\bibfnamefont {J.~B.}\ \bibnamefont {Neaton}}, \ and\
  \bibinfo {author} {\bibfnamefont {K.~M.}\ \bibnamefont {Rabe}},\ }\href@noop
  {} {\bibfield  {journal} {\bibinfo  {journal} {Phys. Rev. B}\ }\textbf
  {\bibinfo {volume} {74}},\ \bibinfo {pages} {094104} (\bibinfo {year}
  {2006})}\BibitemShut {NoStop}%
\bibitem [{\citenamefont {D'Hondt}\ \emph {et~al.}(2008)\citenamefont
  {D'Hondt}, \citenamefont {Abakumov}, \citenamefont {Hadermann}, \citenamefont
  {Kalyuzhnaya}, \citenamefont {Rozova}, \citenamefont {Antipov},\ and\
  \citenamefont {Van~Tendeloo}}]{Dhondt_etal:2008}%
  \BibitemOpen
  \bibfield  {author} {\bibinfo {author} {\bibfnamefont {H.}~\bibnamefont
  {D'Hondt}}, \bibinfo {author} {\bibfnamefont {A.~M.}\ \bibnamefont
  {Abakumov}}, \bibinfo {author} {\bibfnamefont {J.}~\bibnamefont {Hadermann}},
  \bibinfo {author} {\bibfnamefont {A.~S.}\ \bibnamefont {Kalyuzhnaya}},
  \bibinfo {author} {\bibfnamefont {M.~G.}\ \bibnamefont {Rozova}}, \bibinfo
  {author} {\bibfnamefont {E.~V.}\ \bibnamefont {Antipov}}, \ and\ \bibinfo
  {author} {\bibfnamefont {G.~V.}\ \bibnamefont {Van~Tendeloo}},\ }\href@noop
  {} {\bibfield  {journal} {\bibinfo  {journal} {Chem. Mater.}\ }\textbf
  {\bibinfo {volume} {20}},\ \bibinfo {pages} {7188} (\bibinfo {year}
  {2008})}\BibitemShut {NoStop}%
\bibitem [{\citenamefont {Abakumov}\ \emph {et~al.}(2005)\citenamefont
  {Abakumov}, \citenamefont {Kalyuzhnaya}, \citenamefont {Rozova},
  \citenamefont {Antipov}, \citenamefont {J.},\ and\ \citenamefont
  {Van~Tendeloo}}]{Abakumov_etal:2005}%
  \BibitemOpen
  \bibfield  {author} {\bibinfo {author} {\bibfnamefont {A.~M.}\ \bibnamefont
  {Abakumov}}, \bibinfo {author} {\bibfnamefont {A.~S.}\ \bibnamefont
  {Kalyuzhnaya}}, \bibinfo {author} {\bibfnamefont {M.~G.}\ \bibnamefont
  {Rozova}}, \bibinfo {author} {\bibfnamefont {E.~V.}\ \bibnamefont {Antipov}},
  \bibinfo {author} {\bibfnamefont {H.}~\bibnamefont {J.}}, \ and\ \bibinfo
  {author} {\bibfnamefont {G.~V.}\ \bibnamefont {Van~Tendeloo}},\ }\href@noop
  {} {\bibfield  {journal} {\bibinfo  {journal} {Solid State Sci.}\ }\textbf
  {\bibinfo {volume} {7}},\ \bibinfo {pages} {801} (\bibinfo {year}
  {2005})}\BibitemShut {NoStop}%
\bibitem [{\citenamefont {Hadermann}\ \emph {et~al.}(2007)\citenamefont
  {Hadermann}, \citenamefont {Abakumov}, \citenamefont {D'Hondt}, \citenamefont
  {Kalyuzhnaya}, \citenamefont {Rozova}, \citenamefont {Markina}, \citenamefont
  {Mikheev}, \citenamefont {Tristan}, \citenamefont {Klingeler}, \citenamefont
  {B\"{u}chner},\ and\ \citenamefont {Antipov}}]{Hadermann/Abakumov_etal:2007}%
  \BibitemOpen
  \bibfield  {author} {\bibinfo {author} {\bibfnamefont {J.}~\bibnamefont
  {Hadermann}}, \bibinfo {author} {\bibfnamefont {A.~M.}\ \bibnamefont
  {Abakumov}}, \bibinfo {author} {\bibfnamefont {H.}~\bibnamefont {D'Hondt}},
  \bibinfo {author} {\bibfnamefont {A.~S.}\ \bibnamefont {Kalyuzhnaya}},
  \bibinfo {author} {\bibfnamefont {M.~G.}\ \bibnamefont {Rozova}}, \bibinfo
  {author} {\bibfnamefont {M.~M.}\ \bibnamefont {Markina}}, \bibinfo {author}
  {\bibfnamefont {M.~G.}\ \bibnamefont {Mikheev}}, \bibinfo {author}
  {\bibfnamefont {N.}~\bibnamefont {Tristan}}, \bibinfo {author} {\bibfnamefont
  {R.}~\bibnamefont {Klingeler}}, \bibinfo {author} {\bibfnamefont
  {B.}~\bibnamefont {B\"{u}chner}}, \ and\ \bibinfo {author} {\bibfnamefont
  {E.~V.}\ \bibnamefont {Antipov}},\ }\href@noop {} {\bibfield  {journal}
  {\bibinfo  {journal} {J. Mater. Chem.}\ }\textbf {\bibinfo {volume} {17}},\
  \bibinfo {pages} {692} (\bibinfo {year} {2007})}\BibitemShut {NoStop}%
\bibitem [{\citenamefont {Parsons}\ \emph {et~al.}(2009)\citenamefont
  {Parsons}, \citenamefont {D'Hondt}, \citenamefont {Hadermann},\ and\
  \citenamefont {A.}}]{Parsons_etal:2009}%
  \BibitemOpen
  \bibfield  {author} {\bibinfo {author} {\bibfnamefont {T.~G.}\ \bibnamefont
  {Parsons}}, \bibinfo {author} {\bibfnamefont {H.}~\bibnamefont {D'Hondt}},
  \bibinfo {author} {\bibfnamefont {J.}~\bibnamefont {Hadermann}}, \ and\
  \bibinfo {author} {\bibfnamefont {H.~M.}\ \bibnamefont {A.}},\ }\href@noop {}
  {\bibfield  {journal} {\bibinfo  {journal} {Chem. Mater.}\ }\textbf {\bibinfo
  {volume} {21}},\ \bibinfo {pages} {5527} (\bibinfo {year}
  {2009})}\BibitemShut {NoStop}%
\bibitem [{\citenamefont {Arevalo-Lopez}\ and\ \citenamefont
  {Attfield}(2015)}]{Lopez/Attfield_2015}%
  \BibitemOpen
  \bibfield  {author} {\bibinfo {author} {\bibfnamefont {A.~M.}\ \bibnamefont
  {Arevalo-Lopez}}\ and\ \bibinfo {author} {\bibfnamefont {J.~P.}\ \bibnamefont
  {Attfield}},\ }\href@noop {} {\bibfield  {journal} {\bibinfo  {journal}
  {Dalton Trans.}\ } (\bibinfo {year} {2015})}\BibitemShut {NoStop}%
\bibitem [{\citenamefont {Ramezanipour}\ \emph {et~al.}(2010)\citenamefont
  {Ramezanipour}, \citenamefont {Greedan}, \citenamefont {Grosvenor},
  \citenamefont {Britten}, \citenamefont {Cranswick},\ and\ \citenamefont
  {Garlea}}]{Ramezanipour_etal:2010}%
  \BibitemOpen
  \bibfield  {author} {\bibinfo {author} {\bibfnamefont {F.}~\bibnamefont
  {Ramezanipour}}, \bibinfo {author} {\bibfnamefont {J.~E.}\ \bibnamefont
  {Greedan}}, \bibinfo {author} {\bibfnamefont {A.~P.}\ \bibnamefont
  {Grosvenor}}, \bibinfo {author} {\bibfnamefont {J.~F.}\ \bibnamefont
  {Britten}}, \bibinfo {author} {\bibfnamefont {L.~M.~D.}\ \bibnamefont
  {Cranswick}}, \ and\ \bibinfo {author} {\bibfnamefont {V.~O.}\ \bibnamefont
  {Garlea}},\ }\href@noop {} {\bibfield  {journal} {\bibinfo  {journal} {Chem.
  Mater.}\ }\textbf {\bibinfo {volume} {22}},\ \bibinfo {pages} {6008}
  (\bibinfo {year} {2010})}\BibitemShut {NoStop}%
\bibitem [{\citenamefont {Zhang}\ \emph {et~al.}(2014)\citenamefont {Zhang},
  \citenamefont {Zheng}, \citenamefont {Malliakas}, \citenamefont {Allred},
  \citenamefont {Ren}, \citenamefont {Li}, \citenamefont {Han},\ and\
  \citenamefont {Mitchell}}]{Zhang/Mitchell_2014}%
  \BibitemOpen
  \bibfield  {author} {\bibinfo {author} {\bibfnamefont {J.}~\bibnamefont
  {Zhang}}, \bibinfo {author} {\bibfnamefont {H.}~\bibnamefont {Zheng}},
  \bibinfo {author} {\bibfnamefont {C.~D.}\ \bibnamefont {Malliakas}}, \bibinfo
  {author} {\bibfnamefont {J.~M.}\ \bibnamefont {Allred}}, \bibinfo {author}
  {\bibfnamefont {Y.}~\bibnamefont {Ren}}, \bibinfo {author} {\bibfnamefont
  {Q.}~\bibnamefont {Li}}, \bibinfo {author} {\bibfnamefont {T.-H.}\
  \bibnamefont {Han}}, \ and\ \bibinfo {author} {\bibfnamefont {J.~F.}\
  \bibnamefont {Mitchell}},\ }\href@noop {} {\bibfield  {journal} {\bibinfo
  {journal} {Chem. Mater.}\ }\textbf {\bibinfo {volume} {16}},\ \bibinfo
  {pages} {7172} (\bibinfo {year} {2014})}\BibitemShut {NoStop}%
\bibitem [{\citenamefont {Goldschmidt}(1926)}]{Goldschmidt:1926}%
  \BibitemOpen
  \bibfield  {author} {\bibinfo {author} {\bibfnamefont {V.~M.}\ \bibnamefont
  {Goldschmidt}},\ }\href@noop {} {\bibfield  {journal} {\bibinfo  {journal}
  {Naturwissenschaften}\ }\textbf {\bibinfo {volume} {14}},\ \bibinfo {pages}
  {477} (\bibinfo {year} {1926})}\BibitemShut {NoStop}%
\bibitem [{Note1()}]{Note1}%
  \BibitemOpen
  \bibinfo {note} {Although this is generally intended for use in perovskites,
  it is also applicable to perovskite-derived structures such as these
  brownmillerites.}\BibitemShut {Stop}%
\bibitem [{\citenamefont {Shimakawa}\ \emph {et~al.}(2010)\citenamefont
  {Shimakawa}, \citenamefont {Inoue}, \citenamefont {Haruta}, \citenamefont
  {Kawai}, \citenamefont {Matsumoto}, \citenamefont {Sakaiguchi}, \citenamefont
  {Ichikawa}, \citenamefont {Isoda},\ and\ \citenamefont
  {Kurata}}]{Shimakawa_etal:2010}%
  \BibitemOpen
  \bibfield  {author} {\bibinfo {author} {\bibfnamefont {Y.}~\bibnamefont
  {Shimakawa}}, \bibinfo {author} {\bibfnamefont {S.}~\bibnamefont {Inoue}},
  \bibinfo {author} {\bibfnamefont {M.}~\bibnamefont {Haruta}}, \bibinfo
  {author} {\bibfnamefont {M.}~\bibnamefont {Kawai}}, \bibinfo {author}
  {\bibfnamefont {K.}~\bibnamefont {Matsumoto}}, \bibinfo {author}
  {\bibfnamefont {A.}~\bibnamefont {Sakaiguchi}}, \bibinfo {author}
  {\bibfnamefont {N.}~\bibnamefont {Ichikawa}}, \bibinfo {author}
  {\bibfnamefont {S.}~\bibnamefont {Isoda}}, \ and\ \bibinfo {author}
  {\bibfnamefont {H.}~\bibnamefont {Kurata}},\ }\href@noop {} {\bibfield
  {journal} {\bibinfo  {journal} {Cryst. Growth Des.}\ }\textbf {\bibinfo
  {volume} {10}},\ \bibinfo {pages} {4713} (\bibinfo {year}
  {2010})}\BibitemShut {NoStop}%
\bibitem [{\citenamefont {Inoue}\ \emph {et~al.}(2008)\citenamefont {Inoue},
  \citenamefont {Kawai}, \citenamefont {Shimakawa}, \citenamefont {Mizumaki},
  \citenamefont {Kawamura}, \citenamefont {Watanabe}, \citenamefont
  {Tsujimoto}, \citenamefont {Kageyama},\ and\ \citenamefont
  {Yoshimura}}]{Inoue_etal:2008}%
  \BibitemOpen
  \bibfield  {author} {\bibinfo {author} {\bibfnamefont {S.}~\bibnamefont
  {Inoue}}, \bibinfo {author} {\bibfnamefont {M.}~\bibnamefont {Kawai}},
  \bibinfo {author} {\bibfnamefont {Y.}~\bibnamefont {Shimakawa}}, \bibinfo
  {author} {\bibfnamefont {M.}~\bibnamefont {Mizumaki}}, \bibinfo {author}
  {\bibfnamefont {N.}~\bibnamefont {Kawamura}}, \bibinfo {author}
  {\bibfnamefont {T.}~\bibnamefont {Watanabe}}, \bibinfo {author}
  {\bibfnamefont {Y.}~\bibnamefont {Tsujimoto}}, \bibinfo {author}
  {\bibfnamefont {H.}~\bibnamefont {Kageyama}}, \ and\ \bibinfo {author}
  {\bibfnamefont {K.}~\bibnamefont {Yoshimura}},\ }\href@noop {} {\bibfield
  {journal} {\bibinfo  {journal} {Appl. Phys. Lett.}\ }\textbf {\bibinfo
  {volume} {92}},\ \bibinfo {pages} {161911} (\bibinfo {year}
  {2008})}\BibitemShut {NoStop}%
\bibitem [{\citenamefont {Rossell}\ \emph {et~al.}(2004)\citenamefont
  {Rossell}, \citenamefont {Lebedev}, \citenamefont {Van~Tendeloo},
  \citenamefont {Hayashi}, \citenamefont {Terashima},\ and\ \citenamefont
  {Takano}}]{Rossell/Lebedev_etal:2004}%
  \BibitemOpen
  \bibfield  {author} {\bibinfo {author} {\bibfnamefont {M.~D.}\ \bibnamefont
  {Rossell}}, \bibinfo {author} {\bibfnamefont {O.~I.}\ \bibnamefont
  {Lebedev}}, \bibinfo {author} {\bibfnamefont {G.~V.}\ \bibnamefont
  {Van~Tendeloo}}, \bibinfo {author} {\bibfnamefont {N.}~\bibnamefont
  {Hayashi}}, \bibinfo {author} {\bibfnamefont {T.}~\bibnamefont {Terashima}},
  \ and\ \bibinfo {author} {\bibfnamefont {M.}~\bibnamefont {Takano}},\
  }\href@noop {} {\bibfield  {journal} {\bibinfo  {journal} {J.~Appl. Phys.}\
  }\textbf {\bibinfo {volume} {95}},\ \bibinfo {pages} {5145} (\bibinfo {year}
  {2004})}\BibitemShut {NoStop}%
\bibitem [{\citenamefont {Glazer}(1972)}]{Glazer:1972}%
  \BibitemOpen
  \bibfield  {author} {\bibinfo {author} {\bibfnamefont {A.~M.}\ \bibnamefont
  {Glazer}},\ }\href@noop {} {\bibfield  {journal} {\bibinfo  {journal} {Acta.
  Cryst.}\ }\textbf {\bibinfo {volume} {B28}},\ \bibinfo {pages} {3384}
  (\bibinfo {year} {1972})}\BibitemShut {NoStop}%
\bibitem [{\citenamefont {Vailionis}\ \emph {et~al.}(2011)\citenamefont
  {Vailionis}, \citenamefont {Boschker}, \citenamefont {Siemons}, \citenamefont
  {Houwman}, \citenamefont {Blank}, \citenamefont {Rijnder},\ and\
  \citenamefont {Koster}}]{Vailionis_etal:2011}%
  \BibitemOpen
  \bibfield  {author} {\bibinfo {author} {\bibfnamefont {A.}~\bibnamefont
  {Vailionis}}, \bibinfo {author} {\bibfnamefont {H.}~\bibnamefont {Boschker}},
  \bibinfo {author} {\bibfnamefont {W.}~\bibnamefont {Siemons}}, \bibinfo
  {author} {\bibfnamefont {E.~P.}\ \bibnamefont {Houwman}}, \bibinfo {author}
  {\bibfnamefont {D.~H.~A.}\ \bibnamefont {Blank}}, \bibinfo {author}
  {\bibfnamefont {G.}~\bibnamefont {Rijnder}}, \ and\ \bibinfo {author}
  {\bibfnamefont {G.}~\bibnamefont {Koster}},\ }\href@noop {} {\bibfield
  {journal} {\bibinfo  {journal} {Phys. Rev. B}\ }\textbf {\bibinfo {volume}
  {88}},\ \bibinfo {pages} {064101} (\bibinfo {year} {2011})}\BibitemShut
  {NoStop}%
\bibitem [{\citenamefont {Rondinelli}\ and\ \citenamefont
  {Spaldin}(2011)}]{Rondinelli/Spaldin:2011}%
  \BibitemOpen
  \bibfield  {author} {\bibinfo {author} {\bibfnamefont {J.~M.}\ \bibnamefont
  {Rondinelli}}\ and\ \bibinfo {author} {\bibfnamefont {N.~A.}\ \bibnamefont
  {Spaldin}},\ }\href@noop {} {\bibfield  {journal} {\bibinfo  {journal} {Adv.
  Mater.}\ }\textbf {\bibinfo {volume} {23}},\ \bibinfo {pages} {3363}
  (\bibinfo {year} {2011})}\BibitemShut {NoStop}%
\bibitem [{\citenamefont {Rondinelli}\ \emph {et~al.}(2012)\citenamefont
  {Rondinelli}, \citenamefont {May},\ and\ \citenamefont
  {Freeland}}]{Rondinelli/May/Freeland:2012}%
  \BibitemOpen
  \bibfield  {author} {\bibinfo {author} {\bibfnamefont {J.~M.}\ \bibnamefont
  {Rondinelli}}, \bibinfo {author} {\bibfnamefont {S.~J.}\ \bibnamefont {May}},
  \ and\ \bibinfo {author} {\bibfnamefont {J.~W.}\ \bibnamefont {Freeland}},\
  }\href@noop {} {\bibfield  {journal} {\bibinfo  {journal} {MRS Bull.}\
  }\textbf {\bibinfo {volume} {37}},\ \bibinfo {pages} {261} (\bibinfo {year}
  {2012})}\BibitemShut {NoStop}%
\bibitem [{\citenamefont {Johnson-Wilke}\ \emph {et~al.}(2013)\citenamefont
  {Johnson-Wilke}, \citenamefont {Marincel}, \citenamefont {Zhu}, \citenamefont
  {Warusawithana}, \citenamefont {Hatt}, \citenamefont {Sayre}, \citenamefont
  {Delaney}, \citenamefont {Engel-Herber}, \citenamefont {Schlep\"{u}tz},
  \citenamefont {Kim}, \citenamefont {Gopalan}, \citenamefont {Spaldin},
  \citenamefont {Schlom}, \citenamefont {Ryan},\ and\ \citenamefont
  {Trolier-McKinstry}}]{Spaldin/Schlom_etal:2013}%
  \BibitemOpen
  \bibfield  {author} {\bibinfo {author} {\bibfnamefont {R.~L.}\ \bibnamefont
  {Johnson-Wilke}}, \bibinfo {author} {\bibfnamefont {D.}~\bibnamefont
  {Marincel}}, \bibinfo {author} {\bibfnamefont {S.}~\bibnamefont {Zhu}},
  \bibinfo {author} {\bibfnamefont {M.~P.}\ \bibnamefont {Warusawithana}},
  \bibinfo {author} {\bibfnamefont {A.}~\bibnamefont {Hatt}}, \bibinfo {author}
  {\bibfnamefont {J.}~\bibnamefont {Sayre}}, \bibinfo {author} {\bibfnamefont
  {K.~T.}\ \bibnamefont {Delaney}}, \bibinfo {author} {\bibfnamefont
  {R.}~\bibnamefont {Engel-Herber}}, \bibinfo {author} {\bibfnamefont {C.~M.}\
  \bibnamefont {Schlep\"{u}tz}}, \bibinfo {author} {\bibfnamefont {J.-W.}\
  \bibnamefont {Kim}}, \bibinfo {author} {\bibfnamefont {V.}~\bibnamefont
  {Gopalan}}, \bibinfo {author} {\bibfnamefont {N.~A.}\ \bibnamefont
  {Spaldin}}, \bibinfo {author} {\bibfnamefont {D.~G.}\ \bibnamefont {Schlom}},
  \bibinfo {author} {\bibfnamefont {P.~J.}\ \bibnamefont {Ryan}}, \ and\
  \bibinfo {author} {\bibfnamefont {S.}~\bibnamefont {Trolier-McKinstry}},\
  }\href@noop {} {\bibfield  {journal} {\bibinfo  {journal} {Phys. Rev. B}\
  }\textbf {\bibinfo {volume} {88}},\ \bibinfo {pages} {174101} (\bibinfo
  {year} {2013})}\BibitemShut {NoStop}%
\bibitem [{\citenamefont {Eng}\ \emph {et~al.}(2003)\citenamefont {Eng},
  \citenamefont {Barnes}, \citenamefont {Auer},\ and\ \citenamefont
  {Woodward}}]{Eng/Woodward:2003}%
  \BibitemOpen
  \bibfield  {author} {\bibinfo {author} {\bibfnamefont {H.~W.}\ \bibnamefont
  {Eng}}, \bibinfo {author} {\bibfnamefont {P.~W.}\ \bibnamefont {Barnes}},
  \bibinfo {author} {\bibfnamefont {B.~M.}\ \bibnamefont {Auer}}, \ and\
  \bibinfo {author} {\bibfnamefont {P.~M.}\ \bibnamefont {Woodward}},\
  }\href@noop {} {\bibfield  {journal} {\bibinfo  {journal} {J. Solid State
  Chem.}\ }\textbf {\bibinfo {volume} {175}},\ \bibinfo {pages} {94} (\bibinfo
  {year} {2003})}\BibitemShut {NoStop}%
\bibitem [{\citenamefont {Amat}\ \emph {et~al.}(2014)\citenamefont {Amat},
  \citenamefont {Mosconi}, \citenamefont {Ronca}, \citenamefont {Quarti},
  \citenamefont {Umari}, \citenamefont {Nazeeruddin}, \citenamefont
  {Gr\"{a}tael},\ and\ \citenamefont {Angelis}}]{Amat_etal:2014}%
  \BibitemOpen
  \bibfield  {author} {\bibinfo {author} {\bibfnamefont {A.}~\bibnamefont
  {Amat}}, \bibinfo {author} {\bibfnamefont {E.}~\bibnamefont {Mosconi}},
  \bibinfo {author} {\bibfnamefont {E.}~\bibnamefont {Ronca}}, \bibinfo
  {author} {\bibfnamefont {C.}~\bibnamefont {Quarti}}, \bibinfo {author}
  {\bibfnamefont {P.}~\bibnamefont {Umari}}, \bibinfo {author} {\bibfnamefont
  {M.~K.}\ \bibnamefont {Nazeeruddin}}, \bibinfo {author} {\bibfnamefont
  {M.}~\bibnamefont {Gr\"{a}tael}}, \ and\ \bibinfo {author} {\bibfnamefont
  {F.~D.}\ \bibnamefont {Angelis}},\ }\href@noop {} {\bibfield  {journal}
  {\bibinfo  {journal} {Nano Lett.}\ }\textbf {\bibinfo {volume} {14}},\
  \bibinfo {pages} {3608} (\bibinfo {year} {2014})}\BibitemShut {NoStop}%
\bibitem [{\citenamefont {Filip}\ \emph {et~al.}(2014)\citenamefont {Filip},
  \citenamefont {Eperon}, \citenamefont {Snaith},\ and\ \citenamefont
  {Giustino}}]{Filip_etal:2014}%
  \BibitemOpen
  \bibfield  {author} {\bibinfo {author} {\bibfnamefont {M.}~\bibnamefont
  {Filip}}, \bibinfo {author} {\bibfnamefont {G.~E.}\ \bibnamefont {Eperon}},
  \bibinfo {author} {\bibfnamefont {H.~J.}\ \bibnamefont {Snaith}}, \ and\
  \bibinfo {author} {\bibfnamefont {F.}~\bibnamefont {Giustino}},\ }\href@noop
  {} {\bibfield  {journal} {\bibinfo  {journal} {Nature Commun.}\ }\textbf
  {\bibinfo {volume} {5}},\ \bibinfo {pages} {5757} (\bibinfo {year}
  {2014})}\BibitemShut {NoStop}%
\bibitem [{\citenamefont {Aschauer}\ and\ \citenamefont
  {Spaldin}(2014)}]{Aschauer/Spaldin:2014}%
  \BibitemOpen
  \bibfield  {author} {\bibinfo {author} {\bibfnamefont {U.}~\bibnamefont
  {Aschauer}}\ and\ \bibinfo {author} {\bibfnamefont {N.~A.}\ \bibnamefont
  {Spaldin}},\ }\href@noop {} {\bibfield  {journal} {\bibinfo  {journal} {J.
  Phys.: Condens. Matter}\ }\textbf {\bibinfo {volume} {26}},\ \bibinfo {pages}
  {122203} (\bibinfo {year} {2014})}\BibitemShut {NoStop}%
\bibitem [{\citenamefont {Li}\ \emph {et~al.}(2015)\citenamefont {Li},
  \citenamefont {Castelli}, \citenamefont {Thygesen},\ and\ \citenamefont
  {Jacobsen}}]{Li_etal:2015}%
  \BibitemOpen
  \bibfield  {author} {\bibinfo {author} {\bibfnamefont {H.}~\bibnamefont
  {Li}}, \bibinfo {author} {\bibfnamefont {I.~E.}\ \bibnamefont {Castelli}},
  \bibinfo {author} {\bibfnamefont {K.~S.}\ \bibnamefont {Thygesen}}, \ and\
  \bibinfo {author} {\bibfnamefont {K.~W.}\ \bibnamefont {Jacobsen}},\
  }\href@noop {} {\bibfield  {journal} {\bibinfo  {journal} {Phys. Rev. B}\
  }\textbf {\bibinfo {volume} {91}},\ \bibinfo {pages} {045204} (\bibinfo
  {year} {2015})}\BibitemShut {NoStop}%
\bibitem [{\citenamefont {Imada}\ \emph {et~al.}(1998)\citenamefont {Imada},
  \citenamefont {Fujimori},\ and\ \citenamefont
  {Tokura}}]{Imada/Fujimori/Tokura:1998}%
  \BibitemOpen
  \bibfield  {author} {\bibinfo {author} {\bibfnamefont {M.}~\bibnamefont
  {Imada}}, \bibinfo {author} {\bibfnamefont {A.}~\bibnamefont {Fujimori}}, \
  and\ \bibinfo {author} {\bibfnamefont {Y.}~\bibnamefont {Tokura}},\ }\href
  {\doibase 10.1103/RevModPhys.70.1039} {\bibfield  {journal} {\bibinfo
  {journal} {Rev. Mod. Phys.}\ }\textbf {\bibinfo {volume} {70}},\ \bibinfo
  {pages} {1039} (\bibinfo {year} {1998})}\BibitemShut {NoStop}%
\end{thebibliography}%
\bibliographystyle{apsrev4-1}

\end{document}